\documentclass[aps,pra,amssymb,superscriptaddress,10pt,tightenlines]{revtex4-2}

\usepackage[english]{babel}
\usepackage{graphicx}
\usepackage{dcolumn}
\usepackage{bm}
\usepackage{color}
\usepackage{subfigure}
\usepackage[ansinew]{inputenc}
\usepackage{amsfonts}
\usepackage{amsmath}
\usepackage{amssymb}
\usepackage{subfigure}
\usepackage{bbm}
\usepackage{tikz}

\usepackage[colorlinks=true,linkcolor=red,urlcolor=red,citecolor=magenta,pdfusetitle]{hyperref}

\usepackage[braket]{qcircuit}
\usepackage{braket}

\usepackage{mathrsfs}
\usepackage{dsfont}

\def\e{\mathrm{e}}

\def\ii{\mathrm{i}}

\begin{document}



\begin{center}
\title*{\Large{Quantum error mitigation in optimized circuits for particle-density correlations in real-time dynamics of the Schwinger model}}
\end{center}

\author{Domenico Pomarico}%
\affiliation{%
	Dipartimento di Fisica, Universit\`a di Bari, I-70126 Bari, Italy
}%
\affiliation{
Istituto Nazionale di Fisica Nucleare, Sezione di Bari, I-70126 Bari, Italy}

\author{Mahul Pandey}%
\email{mahulpandey@gmail.com}
\affiliation{
	Istituto Nazionale di Fisica Nucleare, Sezione di Bologna, I-40127 Bologna, Italy}

\author{Riccardo Cioli}%
\email{riccardo.cioli3@unibo.it}
\affiliation{
	Istituto Nazionale di Fisica Nucleare, Sezione di Bologna, I-40127 Bologna, Italy}
\affiliation{%
	Dipartimento di Fisica e Astronomia, Universit\`a di Bologna, I-40127 Bologna, Italy
}%

\author{Federico Dell'Anna}%
\affiliation{
	Istituto Nazionale di Fisica Nucleare, Sezione di Bologna, I-40127 Bologna, Italy}
\affiliation{%
	Dipartimento di Fisica e Astronomia, Universit\`a di Bologna, I-40127 Bologna, Italy
}%

\author{Saverio Pascazio}%
\affiliation{%
	Dipartimento  di Fisica, Universit\`a di Bari, I-70126 Bari, Italy
}%
\affiliation{
	Istituto Nazionale di Fisica Nucleare, Sezione di Bari, I-70126 Bari, Italy}

\author{Francesco V. Pepe}%
\affiliation{%
	Dipartimento di Fisica, Universit\`a di Bari, I-70126 Bari, Italy
}%
\affiliation{
	Istituto Nazionale di Fisica Nucleare, Sezione di Bari, I-70126 Bari, Italy}

\author{Paolo Facchi}%
\affiliation{%
	Dipartimento di Fisica, Universit\`a di Bari, I-70126 Bari, Italy
}%
\affiliation{
	Istituto Nazionale di Fisica Nucleare, Sezione di Bari, I-70126 Bari, Italy}

\author{Elisa Ercolessi}%
\affiliation{
	Istituto Nazionale di Fisica Nucleare, Sezione di Bologna, I-40127 Bologna, Italy}
\affiliation{%
	Dipartimento di Fisica e Astronomia, Universit\`a di Bologna, I-40127 Bologna, Italy
}%

\begin{abstract}
	Quantum computing gives direct access to the study of real-time dynamics of quantum many-body systems. In principle, it is possible to directly calculate non-equal-time correlation functions, from which one can detect interesting phenomena, such as the presence of quantum scars or dynamical quantum phase transitions. In practice, these calculations are strongly affected by noise, due to the complexity of the required quantum circuits. As a testbed for the evaluation of real-time evolution of observables and correlations,  the lattice $\mathbb Z_n$ Schwinger model dynamics is considered in one dimension. To control the computational cost, we adopt a quantum-classical strategy that reduces the dimensionality of the system by restricting the dynamics to the Dirac vacuum sector and optimizes the embedding into a qubit model by minimizing the number of three-qubit gates. We derive a digital circuit implementation of time-evolution of particle-density correlation operators and their correlation, comparing results from exact evolution, bare noisy simulations and simulations with different error mitigation techniques. For the evolution of the particle-density operators we also perform runs on a physical IBM quantum device.
\end{abstract}

\maketitle

\pagecolor{white}

\section{Introduction}\label{into}

Investigations of lattice gauge theories constitute an interface among fundamental physics, characterization of quantum many-body systems, and quantum computing implementations by means of properly engineered coding strategies \cite{Banuls2020,dimeglio2023quantum}. The availability of noisy intermediate-scale quantum (NISQ) devices in cloud access platforms opens interesting possibilities from both academic and industrial perspectives. Nonetheless, the operating quantum systems are critically affected by noise, thus preventing current NISQ devices from actually outperforming classical computing capabilities \cite{stoudenmire2023}. This statement holds true in a diversified way in the cases of different hardware setups engineered for quantum computing purposes, each characterized by different advantages in terms of gate fidelity and experimental realization. 

In this article, we will concentrate on the lattice version of quantum electrodynamics (QED), the so-called Schwinger model \cite{schwinger1962}. A rich variety of phenomena emerges driven by the intrinsic nature of lattice QED as a kinetically constrained model, due to the existence of conditions that limit the space of physical states \cite{papic2018, pxp2020, ergo2023}. This characterization leads to the observation of quantum scarring, preserved through linear gauge protection \cite{zeno, zeno2, zeno3, halimeh2022stabilizing} and detected by means of out-of-time-ordered correlators \cite{otoc2023}. Other relevant observables are represented by non-equal-time correlation functions, used to detect non-analyticities that point out dynamical quantum phase transitions (DQPTs) in quantum quenches \cite{mueller2023quantum, dqpt, dqpt2, dqpt_QLM, dqpt_QLM2, osborne2023probing}. In this context, particle density is specifically relevant, as it can represent the most intuitive quantity to describe, in the continuum limit, the decay of an initial Dirac vacuum state after a quench \cite{ionq2, schwinger1951}.

A first attempt to simulate the dynamics of one-dimensional QED was implemented with ion traps \cite{ionq, ionq2}, targeting the observation of pair production starting from the Dirac vacuum. Further developments adopted quantum-classical algorithms in superconducting circuits, based on embedding the dynamics in a specific vacuum sector \cite{ibm_qed}. Larger system sizes are investigated through analog simulators in optical lattices that host ultracold atoms \cite{analog,analog2}. In setups accessible via cloud platforms, noise that affects computation \cite{willsch2024observation,marchegiani2022quasiparticles,connolly2024coexistence,krause2024quasiparticle,yelton2024modeling,fischer2024nonequilibrium,
mcewen2024resisting,iaia2022phonon} can limit the observation of targeted phenomena \cite{entropy, tancara2023quantum}, and error mitigation techniques must be implemented to recover physically meaningful results \cite{mitigation, mitigation2, mitigation3, javanmard2022quantum, farrell2023scalable, hidalgo2023quantum, filippov2023scalabletensornetworkerrormitigation, fischer2024dynamicalsimulationsmanybodyquantum}. 

The novelty of our investigation is twofold: on one hand, it is based on the evaluation of particle density correlation functions by means of specifically engineered circuits exploiting an ancilla qubit, as proposed for retarded Green functions \cite{endo2020, tavernelli2022, tavernellib2022}; on the other hand, to face the increased experimental complexity measured in terms of circuit depth, we propose an algorithm that makes use of a classical-quantum procedure to optimize the embedding of our model into a qubit system. 

The content is as follows.  In Section~\ref{model} we give a brief introduction to the Schwinger model on a 1-dimensional lattice and with a discretized gauge symmetry group, which is mapped into a quantum spin system via a Jordan-Wigner transformation. In Section~\ref{simulation} we adopt a quantum-classical embedding of the considered dynamics, by first restricting the dynamics in the Dirac vacuum sector by means of translation and charge conjugation symmetries and then choosing the optimal permutation of states that minimizes the number of three-qubit interactions in the Hamiltonian. Then in Section~\ref{secCF} we describe the observables we are interested in, namely particle-density operators and their correlation functions, and derive the digital circuits that implement their real-time evolution. Section~\ref{results} presents our results, run for a lattice composed by $N=4$ sites, comparing exact evolution, bare noisy simulations and simulations with error mitigation techniques, such as twirled readout error extinction (T-REx) \cite{trex} and zero noise extrapolation (ZNE) \cite{zne}. For the evolution of the particle-density operators we have also performed runs on a physical IBM device. Finally, we draw our conclusions in Section \ref{conclusion} and collect some details of our calculations in Appendices \ref{secA1}-\ref{secD1}.

\section{Lattice QED in one spatial dimension} \label{model}

Lattice QED in $(1+1)$ dimensions, representing the spatial discretization of the Schwinger model, is a prototypical case of a gauge theory formulated in reduced spatial dimensionality. In this model, gauge degrees of freedom consist of a single (longitudinal) component of the electric field, interacting with spinless fermions of mass $m$ and charge $g$: canonically anticommuting operators $\psi_x,\;  \psi_x^\dagger$ that represent the matter field live on sites $x$, while each edge with length $a$ connecting sites $x$ and $x+1$ hosts the electric field operator $E_{x,x+1}$ and the gauge connection $U_{x,x+1}=\e^{\ii a A_{x,x+1}}$, where $A_{x,x+1}$ is the vector potential, canonically commuting with the electric field. Since we shall consider a finite lattice with $N$ sites, labelled by $x\in\{0,\dots,N-1\}$, we require periodic boundary conditions and identify the fields corresponding to the index $x=N$ to those at the $x=0$ boundary. The system evolution is determined by the Hamiltonian
\begin{equation} \label{Eq::HmJ}
\mathcal{H} = - \frac{\ii J}{2} \sum_{x=0}^{N-1} \left( \psi_x^\dagger U_{x,x+1} \psi_{x+1} - \mbox{H.c.} \right) + m \sum_{x=0}^{N-1} (-1)^x \psi_x^\dagger \psi_x + \frac{g^2}{2 J} \sum_{x=0}^{N-1} E_{x,x+1}^2.
\end{equation}
The first term represents nearest-neighbor hopping of fermions accompanied by the corresponding transformation of the electric field on the involved lattice edge \cite{ionq, ionq2, entropy}  with $J=a^{-1}$, the second term is the staggered (Kogut-Susskind) mass term which is able to solve  the doubling problem associated to lattice discretization \cite{stagg}, while the last term gives the standard  electric contribution to energy. The Dirac vacuum state, which is realized by creating one fermion in each of the negative-mass (odd-$x$) sites,  coincides with the ground state in the limit of infinite mass \cite{dqpt_QLM,dqpt_QLM2}. We notice that  Hamiltonian  (\ref{Eq::HmJ}) is invariant under charge conjugation and translation by two lattice sites, which, for a lattice with even number $N$ of sites  \cite{Ising}, read as:
\begin{align}
    \mathcal{C}_+ =& \ \begin{cases}
        \psi_x \rightarrow (-1)^{x+1} \psi_{x+1}^\dagger, \ \ & \psi_x^\dagger \rightarrow (-1)^{x+1} \psi_{x+1}, \\
        E_{x,x+1} \rightarrow -E_{x+1,x+2}, \ \ & U_{x,x+1} \rightarrow U_{x+1,x+2}^\dagger,
    \end{cases} \\
    \mathcal{T}_2 =& \ \begin{cases}
    \psi_x \rightarrow \psi_{x+2}, \ \ & \psi_x^\dagger \rightarrow \psi_{x+2}^\dagger, \\
    E_{x,x+1} \rightarrow E_{x+2,x+3}, \ \ & U_{x,x+1} \rightarrow U_{x+2,x+3},
    \end{cases} \\
    \lambda(\mathcal{C}_+) =& \ \{+1,-1,+\ii,-\ii \}, \quad \lambda(\mathcal{T}_2) = \{+1, -1\}, \label{eigenvaluesT_2}
\end{align}
where $\lambda$ denotes the spectrum, $\mathcal{C}_+$ stands for a transformation translating matter and electric fields of a lattice spacing (from site $x$ to site $x+1$), transforming particles into antiparticles and vice versa, while changing sign to the electric field. $\mathcal{C}_-$ acts in the opposite direction. We will also set $\mathcal{C}_- = \mathcal{C}_+^\dagger$ with translation by two lattice sites $\mathcal{T}_2 = \mathcal{C}_+^2$.

A further discretization, involving the local gauge degrees of freedom, consists of replacing the original $U(1)$ gauge group with the finite cyclic group $\mathbb{Z}_n$. Thus the Hilbert space associated to a given link becomes $n$-dimensional \cite{Zn}.  In this case \cite{Zn,Ising}:
\begin{enumerate}
\item a convenient basis of the edge space is represented by the electric field eigenstates $\{ \ket{e_k} \}$, satisfying $E_{x,x+1} \ket{e_k} = e_k \ket{e_k}$, with $e_k = \sqrt{\frac{2\pi}{n}} \left(k-\frac{n-1}{2}\right)$ and $k=0,\dots,n-1$;\\
\item the gauge connection $U_{x,x+1}$ acts on this basis as a cyclic permutation: $U_{x,x+1} \ket{e_k} = \ket{e_{k+1}}$ and $U_{x,x+1} \ket{e_{n-1}} = \ket{e_0}$;\\
\item the lattice counterpart of the Gauss law $G_x \ket{\phi} = 0$ which fixes the admissible states $\ket{\phi}$ is given by:
\begin{equation}
G_x = \sqrt{\frac{n}{2\pi}} \left(E_{x,x+1} - E_{x-1,x}\right) - \psi_x^\dagger \psi_x - \frac{(-1)^x - 1}{2},
\end{equation}
that must be valid at all sites $x$ and at any time.
\end{enumerate}

To take on quantum computation on the model Hamiltonian \eqref{Eq::HmJ}, it is convenient to perform the Jordan-Wigner transformation
\begin{equation}
        \psi_x = \sigma^+_x \prod_{\ell=0}^{x-1} (\mathrm{i} Z_{\ell}) , \qquad \text{with } \sigma^{\pm}_x = \frac{X_x \pm \mathrm{i}Y_x}{2}
\end{equation}
where $X_x$, $Y_x$ and $Z_x$ are Pauli matrices acting on the state of site $x$ in the fermion occupation number basis. The transformation maps the matter field into a collection of $N$ spins (qubits), with the advantage of working with operators that commute at different sites. We also rescale the transformed Hamiltonian \cite{ionq, ionq2, entropy} by setting $\xi = J^2 / g^2$, $\mu = m J / g^2$ to finally get the Hamiltonian rewritten as
\begin{equation} \label{Eq::Hximu}
\mathcal{H}(\xi, \mu) = \xi \sum_{x=0}^{N-1} \left( \sigma_x^- U_{x,x+1} \sigma_{x+1}^+ + \mbox{H.c.} \right) - \mu \sum_{x=0}^{N-1} (-1)^x Z_x + \sum_{x=0}^{N-1} E_{x,x+1}^2.
\end{equation} 

In the following we will consider a lattice composed by $N=4$ sites with two particles. We will also discretize the unitary group with $\mathbb Z_3$. 
Appendix \ref{secA1} contains a description  for the gauge invariant subspace and the orbits of the charge conjugation and translation operators for this case, which identify the basis required to explicitly highlight the Hilbert space sectors.

\section{Classical-Quantum embedding} \label{simulation}

In this Section we will present the algorithm we use to simulate the dynamics of the model described by the Hamiltonian \eqref{Eq::Hximu}. To have an efficient protocol, it is necessary to reduce the number of quantum resources needed for the simulation. We do so by adopting a two-step scheme: first we restrict the computation to the subspace containing the Dirac vacuum and then we consider a qubit embedding in which the Hamiltonian contains the smallest number of three-body terms. 

To reach the first goal, we split the Hilbert space by considering the translation by two lattice sites and charge conjugation symmetry sectors, which are labelled by their eigenvalues $T_2=+1$ and $C=\pm1$, $T_2=-1$ and $C=\pm\ii$ listed in Eq. \eqref{eigenvaluesT_2}. It is thus possible to partition the physical Hilbert space in four sectors, such that the Hamiltonian \eqref{Eq::Hximu} is expressed as $U \mathcal{H} U^\dagger = \bigoplus_{T_2,C} \mathcal{H}^{(T_2,C)}$. We exploit the block structure described in Appendix \ref{secA1} to focus on the $(+,+)$ sector, containing the Dirac vacuum. The restriction of the Hamiltonian to this seven-dimensional subspace reads
\begin{equation} \label{Eq::H++}
    \mathcal{H}^{(+,+)} =  \begin{pmatrix}
         -2\mu & \xi & & & & & \\
         \xi & \frac{\pi}{3} & \frac{\xi}{\sqrt{2}} & & & & \\
         & \frac{\xi}{\sqrt{2}} & 2\mu + \frac{2\pi}{3} & \frac{\xi}{\sqrt{2}} & & & \\
         & & \frac{\xi}{\sqrt{2}} & \pi & \frac{\xi}{\sqrt{2}} & & \\
         & & & \frac{\xi}{\sqrt{2}} & -2\mu + \frac{4\pi}{3} & \frac{\xi}{\sqrt{2}} & \\
         & & & & \frac{\xi}{\sqrt{2}} & \frac{4\pi}{3} & \xi \\
         & & & & & \xi & 2\mu + \frac{4\pi}{3}
    \end{pmatrix},
\end{equation}
according to the ordered basis given in Appendix \ref{secA1}. 

As for the embedding, we notice that seven states can be described by means of three qubits, the elementary approach relying on adding a last column and row filled with zeros to the Hamiltonian matrix to keep the eighth unphysical state idle.  The direct approach consists in identifying the first seven states of a three qubit system with the seven states \eqref{A1}- \eqref{A7} listed in Appendix~\ref{secA1}: doing so the Hamiltonian (\ref{Eq::H++}) can be written as
\begin{equation} \label{Eq::Hnoperm} 
    \widetilde{\mathcal{H}}^{(+,+)} = \frac{1}{8} \sum_{i,j,k=0}^3 \mathrm{Tr} \left\{ \sigma_i \otimes \sigma_j \otimes \sigma_k \widetilde{\mathcal{H}}^{(+,+)}\right\} \sigma_i \otimes \sigma_j \otimes \sigma_k = \sum_{i,j,k=0}^3 c_{(i,j,k)} \sigma_i \otimes \sigma_j \otimes \sigma_k ,
\end{equation}
where $\sigma_0$ is the $2\times 2$ identity matrix and $\sigma_j$ with $j\in\{1,2,3\}$ are the Pauli matrices. A straightforward but lengthy calculation shows that such decomposition accounts for 19 non-zero terms, of which 7 correspond to a product of 3 Pauli matrices. In the implementation of the Trotter evolution for $U(t)=\mathrm{exp}\left(-\ii \widetilde{\mathcal{H}}^{(+,+)} t \right)$, the latter terms require a higher number of $CNOT$ gates. Thus, in order to reduce the computational time, we aim at finding the embedding in which the Hamiltonian contains the smallest number of three-body terms, i.e. we want to minimize  the number of coefficients $c_{(i,j,k)}$ with $i,j,k \in\{1,2,3\}$ by considering all possibile permutations $\pi\in S_8$ of the 8 basis states. For the chosen problem, we find that the optimal permutation is $\pi_o = (7,6,1,2,4,5,8,3)$, leading to a Hamiltonian $\widetilde{\mathcal{H}}^{(+,+)}_{\pi_o(\ell),\pi_o(m)}$ which contains 15 nonzero terms, of which only 3 are triple products of Pauli matrices gates. In Appendix~\ref{secB1} we list all coefficients for both the simple and the optimal embedding in the case we consider here, while in Appendix~\ref{secBB1} we propose an algorithm to find such an optimal permutation in the general case.  Finally, to reduce the usage of $SWAP$ gates, circuit optimization can also rely on a properly chosen topology of superconducting qubits connections  \cite{entropy}.

\section{Digital circuit for unitary evolution and Green functions} \label{secCF}

\begin{figure}
    \centering
    \includegraphics[width = \linewidth]{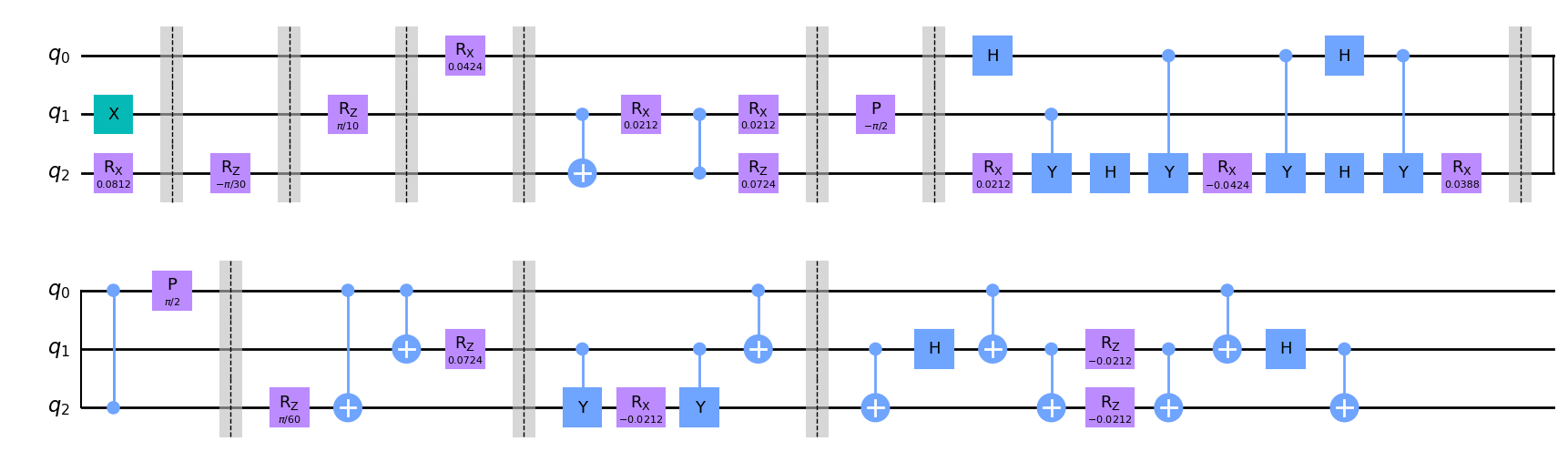}
    \caption{Trotter step circuit for the permutation $\pi_o=(7,6,1,2,4,5,8,3)$ and $\Delta t = 0.1$. $\xi = 0.6$, $\mu=0.1$, with the initial spin flip required to map $\ket{\mathrm{vac}}_i \rightarrow \ket{\mathrm{vac}}_{\pi_o(i)}$.}
    \label{fig:4sites_results_best}
\end{figure}

Once we have obtained the optimal decomposition for the Hamiltonian, the time-evolution operator $U(t)$ is implemented via a Trotter decomposition as described in Appendix~\ref{secB2}. Each Trotter step is therefore represented by the circuit shown in Fig. \ref{fig:4sites_results_best}.

Let us look now at observables of interest. The ones we want to consider are the time evolution of particle-density and its associated correlation functions. The formal expression of the former is \cite{ionq2}
\begin{equation}
    \nu = \frac{1}{N}\sum_x (-1)^x \psi_x^\dagger \psi_x,
\end{equation}
admitting the sector decomposition $U \nu U^\dagger = \bigoplus_{T_2,C} \nu^{(T_2,C)}$, once we restrict it to the physical subspace. In particular, in the $(+,+)$ sector, we have
\begin{equation} \label{Eq::num}
    \widetilde{\nu}^{(+,+)} = \sum_{i,j,k=0}^3 d_{(i,j,k)} \sigma_i \otimes \sigma_j \otimes \sigma_k,
\end{equation}
with a number of nonvanishing coefficients that depends on the specific permutation we choose for the embedding. In Appendix~\ref{secB1} we will list the nonvanishing coefficients for the optimal permutation described above and give the explicit expression \eqref{nuespl} that we will use in the following calculations.

The real-time correlation function is defined as
\begin{equation}
    G(t,s) = G^<(t,s) - G^>(t,s) = - \ii \Theta(t-s) \braket{\mathrm{vac} | \{ \widetilde{\nu}^{(+,+)}(t), \widetilde{\nu}^{(+,+)}(s) \} | \mathrm{vac}},
\end{equation}
where, for any two operators  $\{ A, B \} = AB + BA$ and $\Theta(t)$ is the Heaviside step function.  We can choose two arbitrary times $s, t$ since the correlation function vanishes for $s=0$ because of $\widetilde{\nu}^{(+,+)}(0)\ket{\mathrm{vac}}=0$. $G(t,s)$ is composed of two terms, named the lesser $G^<(t,s)$ and the greater term $G^>(t,s)$. The former quantifies for an initial time $s$ how long the particle-density property persists in the future, while the latter measures this memory effect backward in time.

Here we will focus on the lesser term in order to achieve a trade-off between hardware complexity and physical interpretability. In this way it is possible to emphasize error mitigation outputs for circuits highly affected by controlled double qubits gates, without superposing a too high number of terms which amplifies the sum of independent decoherence effects:
\begin{equation} \label{Eq::green2times}
    G^<(t,s) = - \ii \Theta(t-s) \braket{\mathrm{vac} | \widetilde{\nu}^{(+,+)}(t) \widetilde{\nu}^{(+,+)}(s) | \mathrm{vac}}, 
\end{equation}
which, using the definition of $P_\alpha$ in Eq. \eqref{nuespl} and the explicit value of the coefficient $d_{(0,0,0)}=7/16$, can be rewritten as
\begin{align} \label{Eq::green_global_4sites}
    G^< (t,s) = -\ii \Theta(t-s) & \left[ \frac{7}{16} \left( \braket{\mathrm{vac} | \widetilde{\nu}^{(+,+)}(t) | vac} + \braket{vac | \widetilde{\nu}^{(+,+)}(s) |\mathrm{vac} } \right)  - \frac{49}{256} \right. \nonumber \\
    & \quad \left. + \sum_{\alpha,\beta = 1}^7 \braket{\mathrm{vac} | U^\dagger (t) P_\alpha U(t-s) P_\beta U(s) | \mathrm{vac}} \right].
\end{align}
The first two terms of this expression are the expectation values of the density operator evaluated on the time-evolved vacuum state and can be evaluated using the Trotter decomposition of the evolution operator and the circuit of Fig. \ref{fig:4sites_results_best}. The last term can be instead calculated via the circuit  \cite{tavernelli2022}
\begin{equation} \label{circuit}
 \Qcircuit @C=1em @R=.7em {
	\lstick{\ket{0}} & \gate{H} & \gate{\phi} & \gate{X} & \ctrl{1} & \gate{X} & \ctrl{1} & \gate{H} & \meter \\
	\lstick{\ket{\mathrm{vac}}}  & \qw & \qw &  \gate{U(s)}  & \gate{P_\beta} & \gate{U(t-s)} & \gate{P_\alpha} &  \qw  & \qw
}
\end{equation}
which uses a measurement over an ancilla qubit $A$ (initialized in the state $|0\rangle$). Indeed, denoting with $R$ the register encoding the system (initialized in the vacuum state $|\mathrm{vac}\rangle$) and with $\varrho_{\mathrm{out}}$ the output state of the total circuit, it is not difficult to prove that the measurement on the ancilla yields
\begin{equation} \label{Eq::circuit_green}
    \mathrm{Tr}_{AR}\left\{ (Z \otimes \mathds{1}) \varrho_{\mathrm{out}} \right\} = \mathrm{Re} \left\{ \e^{-\ii \phi} \braket{\mathrm{vac}| U^\dagger(t) P_\alpha U(t-s) P_\beta U(s) |\mathrm{vac}} \right\}.
\end{equation}
In Appendix \ref{secC1} we will discuss how this circuit can be realized by means of a Mach-Zender interferometer \cite{vedral2000,vedral2003}.

\section{Implementation on IBM Quantum platform}\label{results}

The algorithm described in the previous section has been implemented on the IBM Quantum platform \cite{ibm_quantum}, more specifically both on the device \texttt{ibmq\_quito} and by means of noise models available in the Python package \texttt{qiskit}, with error rates updated from the aforementioned device. To limit noise sources in state preparation, gates and measurements, error mitigation techniques have been applied in both cases using tools of the package \texttt{qiskit-ibm-runtime}, as discussed in Appendix \ref{secD1}. In the following we present the results for the problem of $N=4$ sites.

\begin{figure}[t!]
    \centering
    \includegraphics[width = 0.9\linewidth]{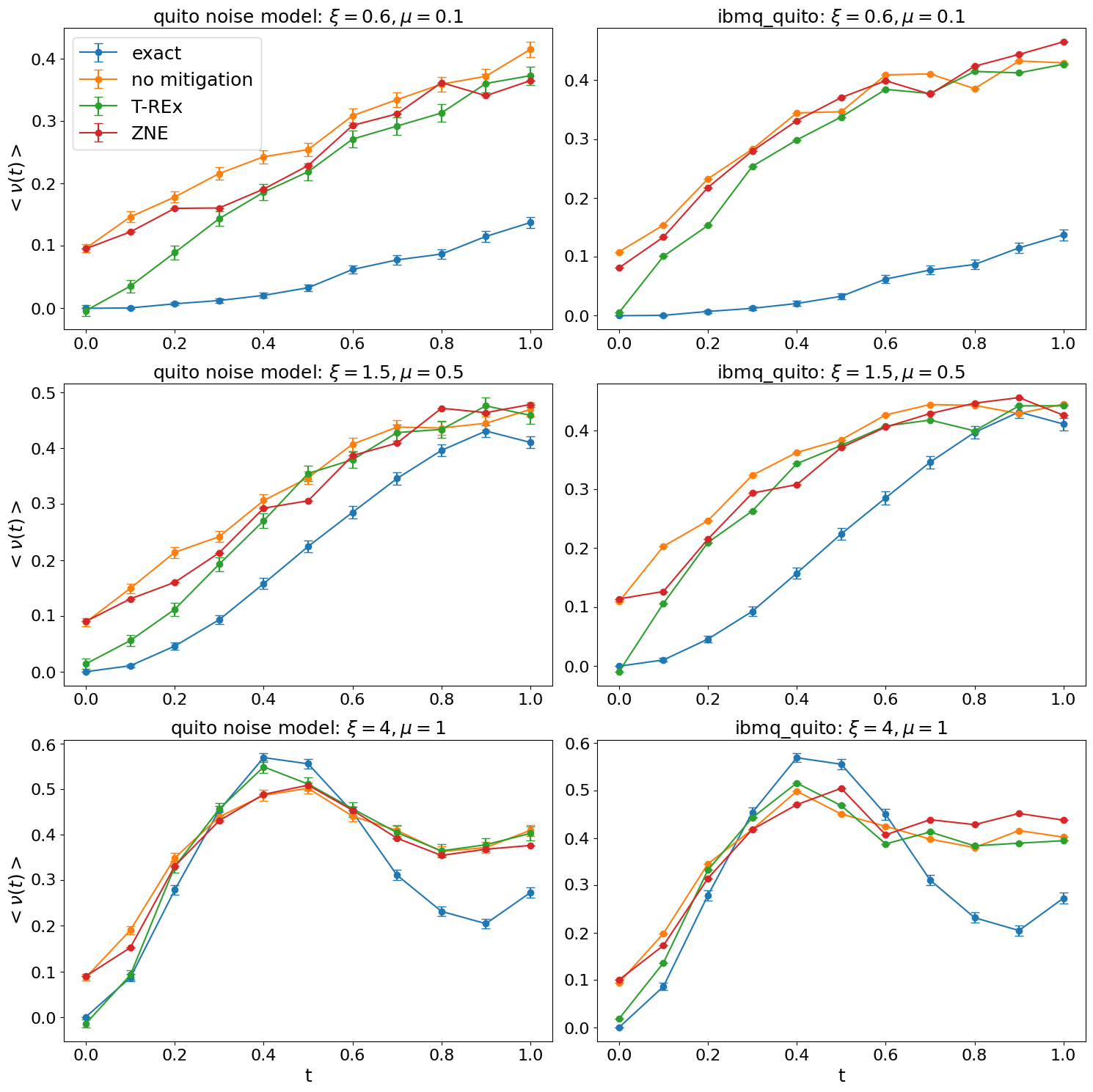}
    \caption{Results for the chosen optimal permutation $\pi=(7,6,1,2,4,5,8,3)$ and Trotter step $\Delta t = 0.1$. Panels in left column are referred to \texttt{ibmq\_quito} noise model, while those in the right one to the device \texttt{ibmq\_quito} measurements.The three rows correspond to the three different coupling regimes $(\xi, \mu) = \{(0.6, 0.1), (1.5, 0.5), (4, 1)\}$ from top to bottom. From top to bottom: strong to weak coupling.}
    \label{fig:4sites_density}
\end{figure}

The real-time evolution of the particle density \eqref{Eq::num} is shown in Fig. \ref{fig:4sites_density}, with the first column showing the noise-simulated data and the second column giving the output of the actual computation on the device.  Since the initial state corresponds to the Dirac vacuum, we can interpret the considered simulation as a quenched dynamics starting from the ground state for $\mu \rightarrow \infty$ in \eqref{Eq::Hximu}, followed by an evolution generated for finite $\mu$ and $\xi$. We call strong (weak) coupling regime a condition in the free parameters space $(\xi, \mu)$ that causes a slowly (rapidly) deviating evolution from the initial Dirac vacuum. The three rows of Fig. \ref{fig:4sites_density} correspond to three different coupling regimes that go from strong to weak coupling, that is, $(\xi, \mu) = \{(0.6, 0.1), (1.5, 0.5), (4, 1)\}$. In each graph, we compare four curves: the exact noiseless simulation (in blue), the bare results of noisy simulation/real results (in orange) and the results processed via two error mitigation techniques, T-REx (in green) and ZNE (in red).

The ZNE mitigation is characterized by an almost identical curve with the unmitigated case, while the T-REx mitigation works better during the first time steps, when noise is mainly dominated by readout errors. We observe that the behavior of noisy and mitigated curves are more affected in the strong coupling case. However, in all cases for longer times the expectation value approximates the maximally mixed condition $\braket{ \widetilde{\nu}^{(+,+)} (t)} = \frac{7}{16}$, showing that noise destroys any coherence. Also, for the real device \texttt{ibmq\_quito}, the strong and intermediate couplings converge faster towards the maximally mixed condition than the corresponding noise model, probably because of the missing correlated noise in the assumed gate stand-alone hypothesis \cite{entropy}, which neglects for example the noise propagation in subsequent applications of double qubit gates sharing a circuit line. The resilient behavior of the weak coupling regime may be due to the presence of the $\xi$ parameter in all double and triple qubits gates presented in Appendix \ref{secB1}. 

\begin{figure}[t!]
\centering
    \subfigure[]{\includegraphics[width=0.45\linewidth]{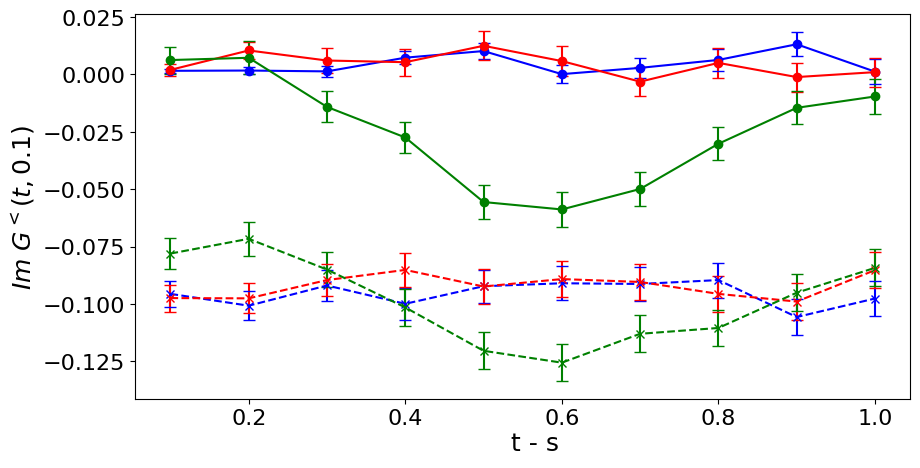}}
    \hfill
    \subfigure[]{\includegraphics[width=0.45\linewidth]{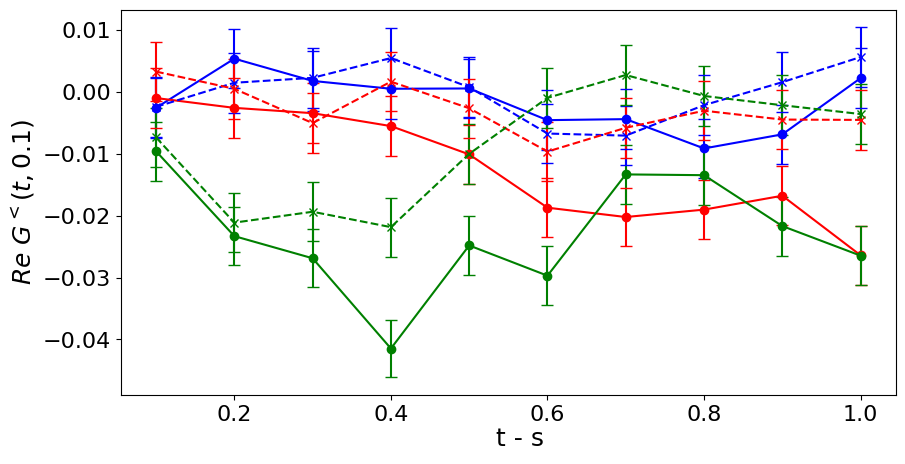}}\\
    \subfigure[]{\includegraphics[width=0.45\linewidth]{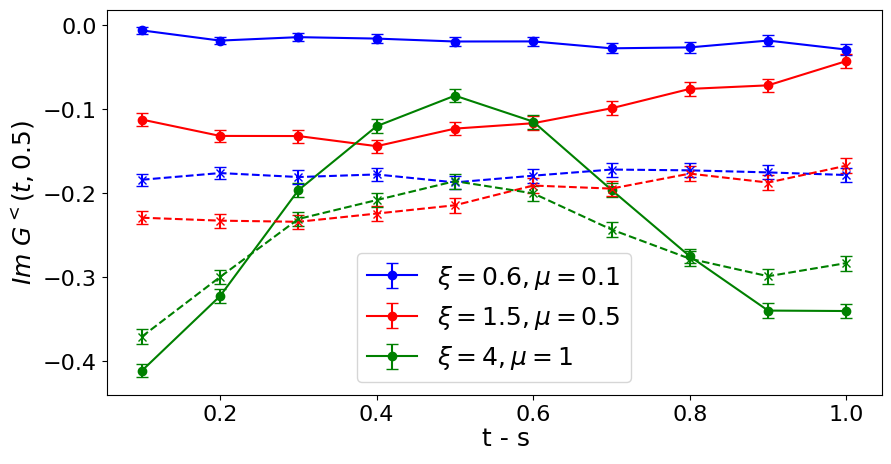}}
    \hfill
    \subfigure[]{\includegraphics[width=0.45\linewidth]{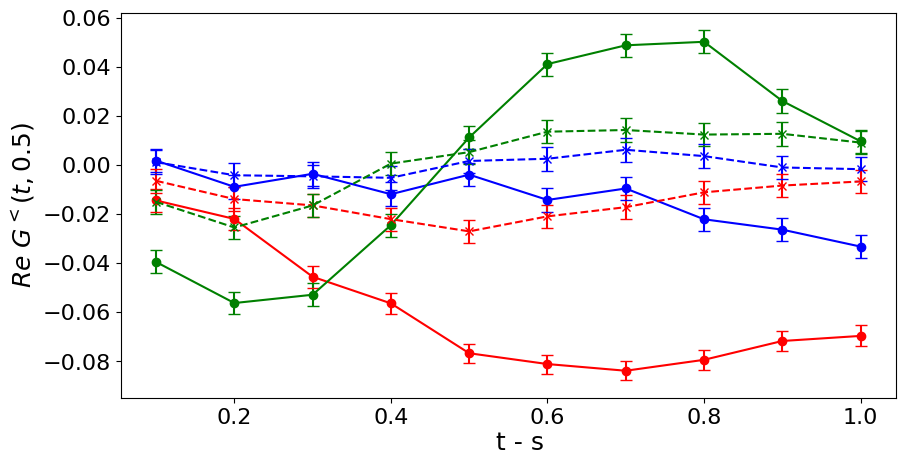}}
    \caption{Bare noise simulation. Lesser contribution to correlation functions of particle-densities $\widetilde{\nu}^{(+,+)}(t)$ and $\widetilde{\nu}^{(+,+)}(s)$ for a non-vanishing time $s$: $s=0.1$ in (a-b), $s=0.5$ in (c-d). Circuits are executed in \texttt{qasm\_simulator}. Solid lines correspond to noiseless Trotter evolution with error bars caused by the finite number of shots, while dashed lines include noise models. Different colors refer to different coupling regimes (see panel c).}
    \label{fig:noisy}
\end{figure}

\begin{figure}[t!]
    \centering
    \subfigure[]{\includegraphics[width=0.45\linewidth]{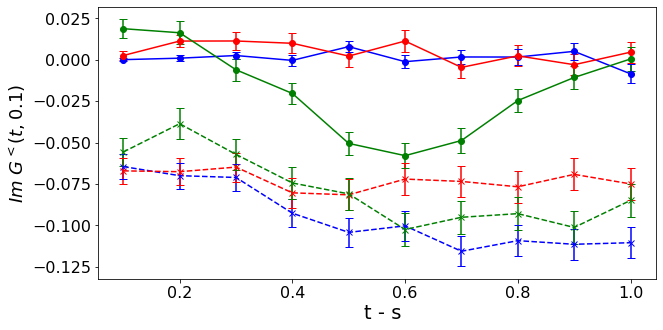}}
    \hfill
    \subfigure[]{\includegraphics[width=0.45\linewidth]{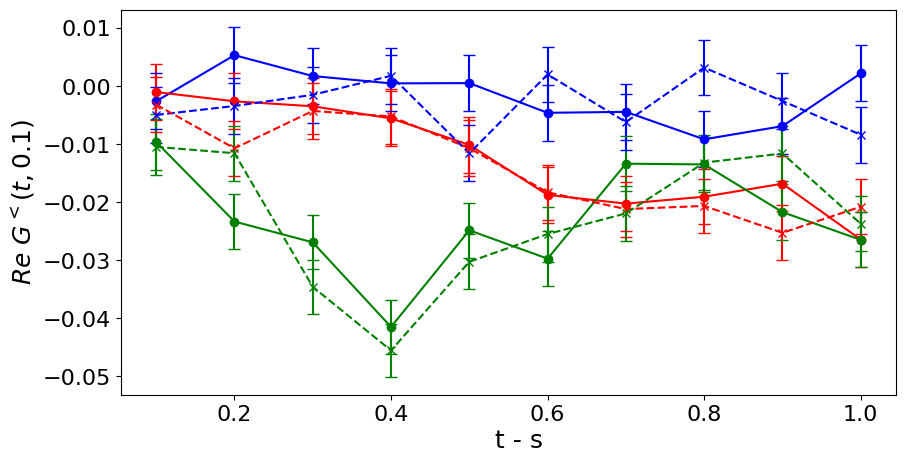}}\\
    \subfigure[]{\includegraphics[width=0.45\linewidth]{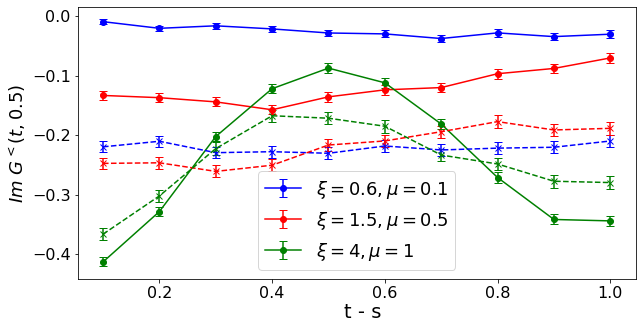}}
    \hfill
    \subfigure[]{\includegraphics[width=0.45\linewidth]{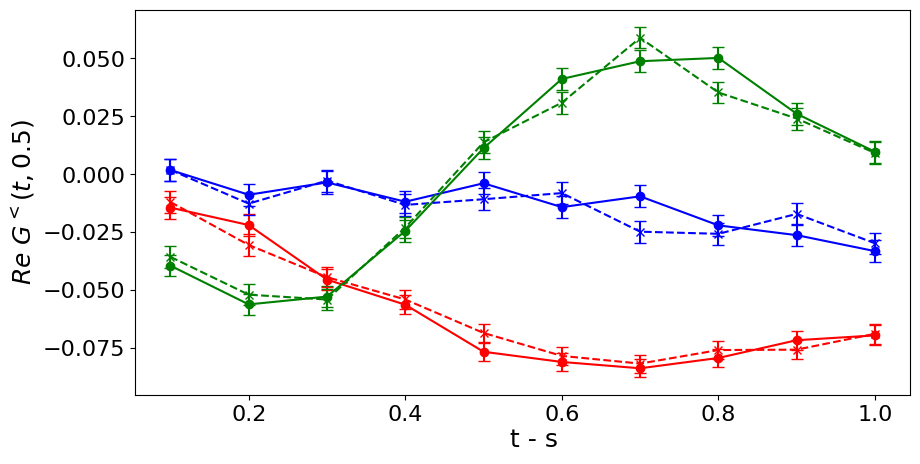}}
    \caption{Simulations processed via T-REx mitigation.  Lesser contribution to correlation functions of particle-densities $\widetilde{\nu}^{(+,+)}(t)$ and $\widetilde{\nu}^{(+,+)}(s)$ for a non-vanishing time $s$: $s=0.1$ in (a-b), $s=0.5$ in (c-d). Circuits are executed in \texttt{qasm\_simulator}. Solid lines correspond to noiseless Trotter evolution with error bars caused by the finite number of shots, while dashed lines include noise models. Different colors refer to different coupling regimes (see panel c).} \label{fig:trex}
\end{figure}

\begin{figure}[t!]
    \centering
    \subfigure[]{\includegraphics[width=0.45\linewidth]{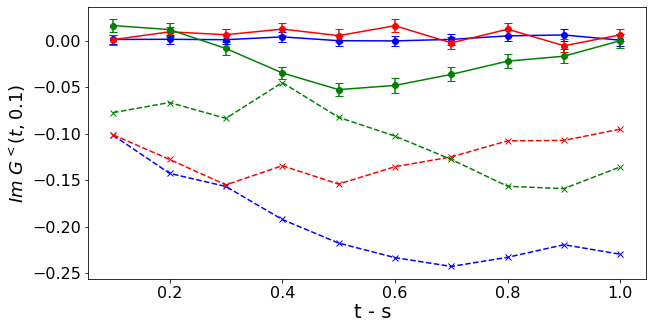}}
    \hfill
    \subfigure[]{\includegraphics[width=0.45\linewidth]{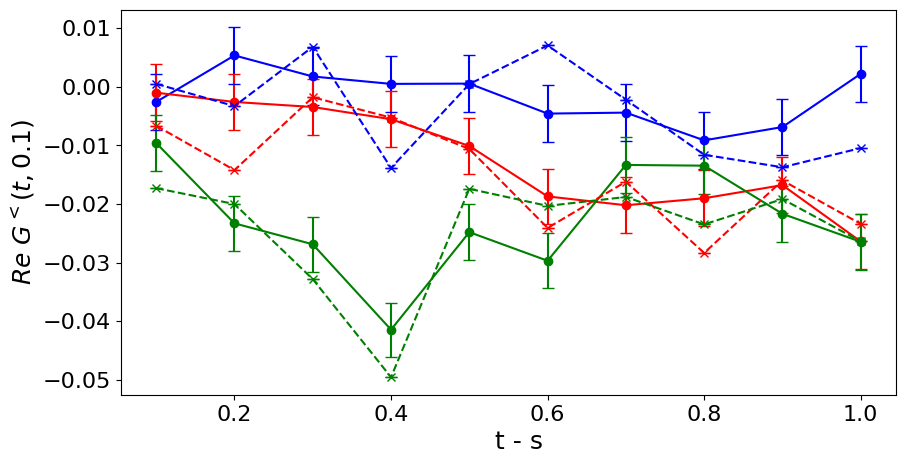}}\\
    \subfigure[]{\includegraphics[width=0.45\linewidth]{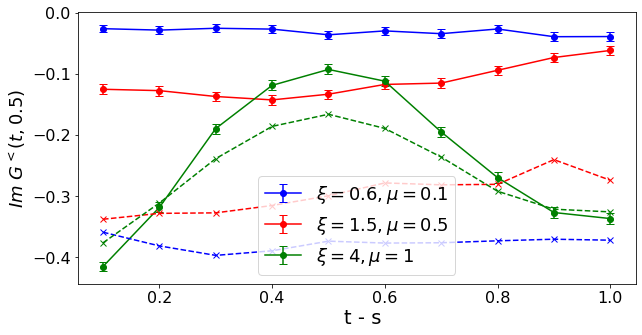}}
    \hfill
    \subfigure[]{\includegraphics[width=0.45\linewidth]{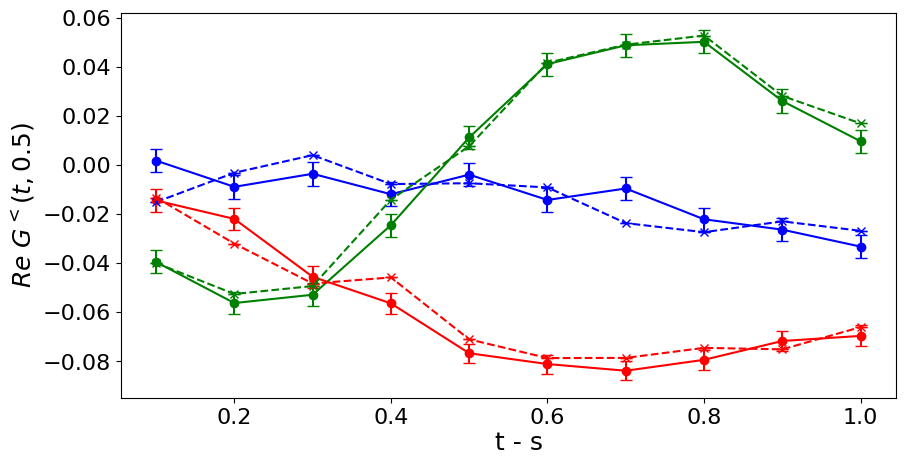}}
    \caption{Simulations processed via ZNE mitigation. Lesser contribution to correlation functions of particle-densities $\widetilde{\nu}^{(+,+)}(t)$ and $\widetilde{\nu}^{(+,+)}(s)$ for a non-vanishing time $s$: $s=0.1$ in (a-b), $s=0.5$ in (c-d). Circuits are executed in \texttt{qasm\_simulator}. Solid lines correspond to noiseless Trotter evolution with error bars caused by the finite number of shots, while dashed lines include noise models. Different colors refer to different coupling regimes (see panel c).} \label{fig:zne}
\end{figure}

The real-time correlation function  $G^<(t,s)$ can be evaluated from Eq. \eqref{Eq::green_global_4sites} only with the aid of the noise models because of the required high circuits depth, implying an execution time longer than the dephasing time of superconducting circuits. To do so, we use the circuit in Eq. \eqref{circuit} by collecting measurements for any pair of Pauli strings operators $P_\alpha P_\beta$ ($\alpha,\beta=1,\cdots,7$) in Eq. \eqref{Eq::green_global_4sites} and by setting the phase $\phi = 0 \ (\frac{\pi}{2})$ to select the imaginary (real) part of lesser contribution. In Fig. \ref{fig:noisy} we collect our results for bare noisy simulations (dashed lines) to be compared to the exact noiseless simulation (solid lines). To interpret the results, we notice that, since particle-densities $\widetilde{\nu}^{(+,+)}(t)$ and $\widetilde{\nu}^{(+,+)}(s)$ are Hermitian observables, they contribute only to the imaginary part of lesser terms, while the 49 terms in Eq. \eqref{Eq::green_global_4sites} related to Pauli strings contribute to both real and imaginary parts. 
This might explain the offset between exact results and bare noise simulations we observe in the imaginary part and not in the real one. Indeed, we can conclude that this offset is caused by the higher error rate of the density $\braket{\widetilde{\nu}^{(+,+)}(t)}$, as observed in Fig. \ref{fig:4sites_density}. On the other hand, the  49 Pauli strings pairs, that might contribute with opposite signs, yield a lower deviation. This offset is almost identical for the imaginary part in panels (a), (c) for any coupling regime, because of the same readout errors and a negligible value of particle-density for low values of $s$. For higher values of $s$ the deviation decreases in stronger couplings, even if damped oscillations with respect to the noiseless case are observed.

In Fig. \ref{fig:trex} and \ref{fig:zne} we collect our results for bare noisy simulation data corrected via T-REx and ZNE mitigation respectively.

As for the former, we notice that real parts of lesser contributions shown in panels (b), (d) are not distinguishable from the noiseless case, thus revealing a significant improvement with respect to the unmitigated case.  Instead imaginary parts in panels (a),(c) are almost unchanged, because the dominant contribution comes from $\widetilde{\nu}^{(+,+)}(t)$ and $\widetilde{\nu}^{(+,+)}(s)$ which coincides in the noisy and mitigated scenario. The weak coupling regime oscillation is well captured with respect to its period in the considered time window, but the convergence towards the maximally mixed state causes amplitude dampings already observed in Fig. \ref{fig:noisy}.

Similar conclusions can be drawn concerning the real part of of lesser contributions corrected by means of ZNE mitigation techniques, as it is shown in panels (b), (d) of Fig. \ref{fig:zne} that show a significantly improved behavior. Instead a non negligible worsening emerges in imaginary parts as shown in panels (a),(c), with the exception of the weak coupling oscillation in panel (c), which is almost unchanged. These behaviors reveal that the circuit folding procedure used in ZNE and discussed in Appendix \ref{secD1} yields a non trivial noise scaling for the considered evaluation of the correlation function imaginary part. For sufficiently low values assumed by the correlation function, errors behave as a small perturbation and are efficiently taken under control by mitigation procedures. Such effects are instead washed out for higher values of the considered function, with a correspondingly increasing order of magnitude for errors.

In general, the measured offset in imaginary parts is caused by an increasing number of contributions driven by mixed states, as explained in Appendix \ref{secC1}, which bring additional contributions with respect to the targeted pure initial Dirac vacuum. A more precise description of output states requires an extension in terms of channels corresponding to gates, in order to target the generation of correlated noise \cite{entropy}, but this goes beyond the scope of this paper. 


\section{Conclusion and outlook} \label{conclusion}

We considered real-time dynamics of the Schwinger model on a periodic lattice in $(1+1)$ dimensions, implemented on IBM Quantum \cite{ibm_quantum} real devices and examined via different noise models. We proposed a quantum-classical approach to minimize the required computational cost: first we used translation by two lattice sites and charge conjugation symmetries to restrict the dynamics to the sector of the Dirac vacuum, then we reduced the number of triple-qubits gates involved in the Trotter evolution by choosing the optimal permutation of states in the qubits embedding. We have examined first the real-time evolution of the particle-density operator in different coupling regimes, showing a resilient behavior with respect to noise in the weak coupling case. Then we calculated time-dependent correlation functions, which are very much affected by noise. Our results prove that error mitigation works properly for small variations of measured quantities. 

The evaluation of the Green function is of fundamental importance for, e.g., the detection of scarring phenomena via the knowledge of retarded Green functions in Fourier space to estimate spectral density peaks \cite{tavernellib2022} corresponding to equally spaced towers \cite{papic2018, ergo2023}. Our analysis can shed some light to understand to what extent decoherence and noise resiliency might affect the analysis of errors occurrence in different  coupling regimes  \cite{dynamics}.

\section{Acknowledgements}

This research is partially funded  by INFN (project ``QUANTUM") and we acknowledge financial
support from the National Centre for HPC, Big Data, and Quantum Computing (Spoke 10,
CN00000013). F.D. and E.E. also acknowledge financial support from the 2022-PRIN Project ``Hybrid algorithms 
for quantum simulators".  D.P., S.P., F.V.P. and P.F. acknowledge support from the Italian funding within the ``Budget MUR - Dipartimenti di Eccellenza 2023-2027" - Quantum Sensing and Modelling for One-Health (QuaSiModO). P.F. acknowledges support from the Italian National Group of Mathematical Physics (GNFM- INdAM) and from PNRR MUR project CN00000013-``Italian National Centre on HPC, Big Data and Quantum Computing". S.P. and F.V.P. acknowledge support from PNRR MUR project PE0000023-NQSTI.

\appendix
\renewcommand\thefigure{\thesection.\arabic{figure}}

\section[\appendixname~\thesection]{Basis of the Gauge invariant subspace \label{secA1}}

In the following we will consider a lattice of $N=4$ links, with two particles, and discretize the unitary group with $\mathbb Z_3$. A state is labeled with the values $e_k = +,0,-$ of the electric field on the four links: $|e_0 \, e_1 \, e_2\, e_3 \rangle$. Out of the possible $3^4$ configurations, only $18$ are gauge invariant, yielding a basis for the physical subspace. The latter can be split into disjoint orbits of the charge conjugation and translation operators.  We notice that  in this particular case \cite{ibm_qed}: $\mathcal{T}_2=\mathcal{C}_+^2=\mathcal{C}_-^2$, so that in order to classify orbits we can simply look at the action of  the charge conjugation operator ${\cal C}_+$.

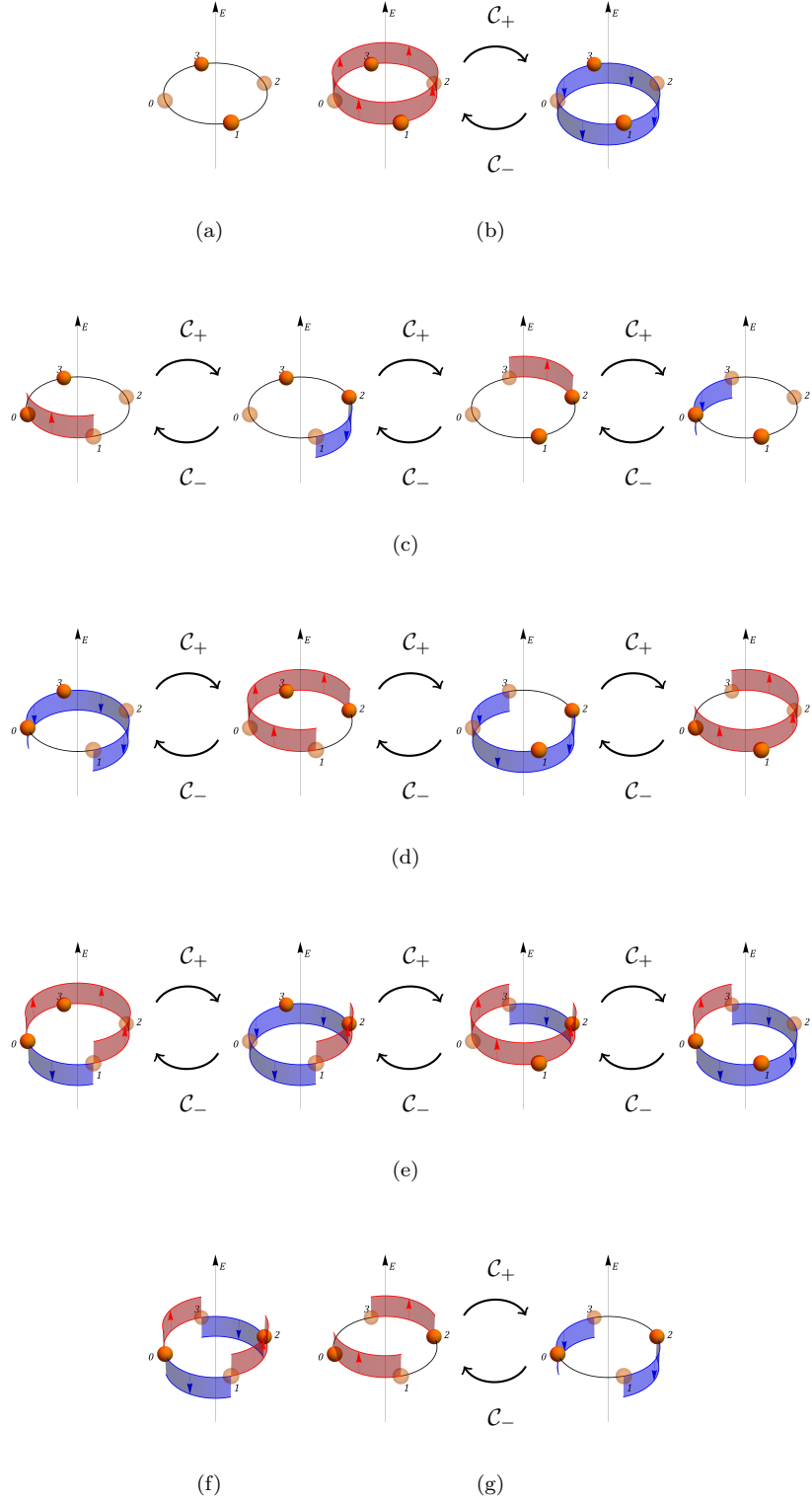
\begin{figure}[t!]
    \centering
    \begin{tabular}{c}
    \vspace{0.1cm}
    \begin{tabular}{cc}
        \subfigure[]{\includegraphics[width=.12\linewidth]{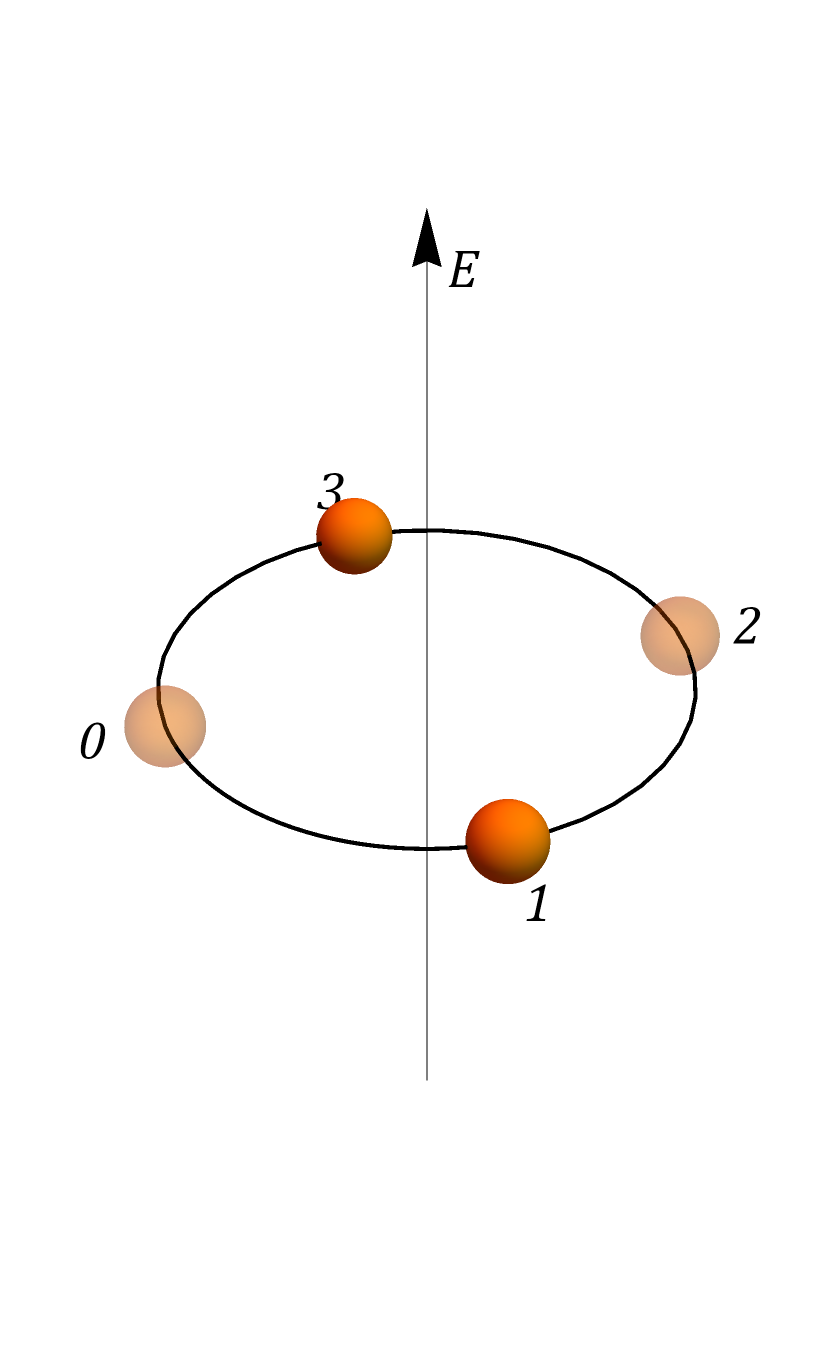}} & \subfigure[]{\begin{tikzpicture}
\node[inner sep=0pt] (russell) at (0,0)
    {\includegraphics[width=.12\linewidth]{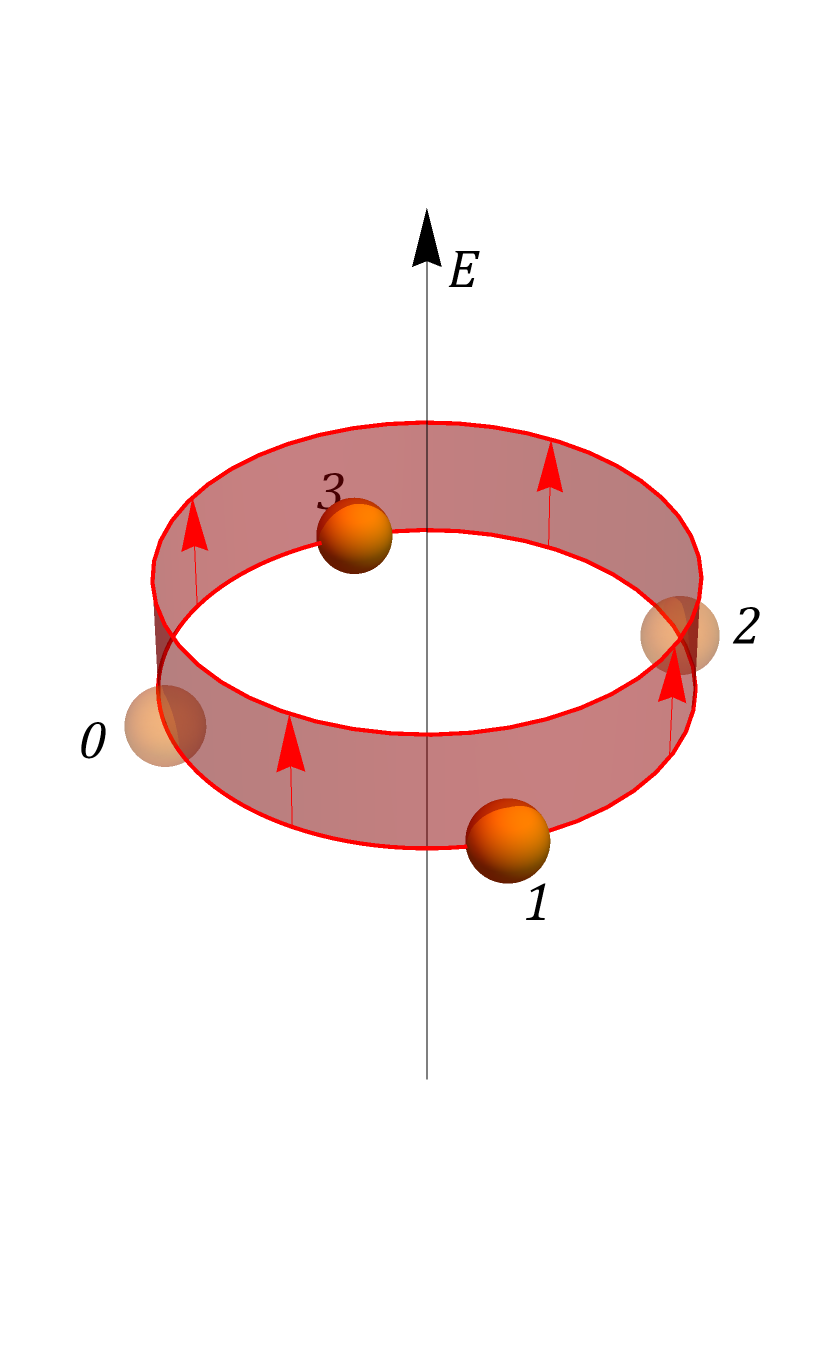}};
\node[inner sep=0pt] (whitehead) at (3,0) {\includegraphics[width=.12\linewidth]{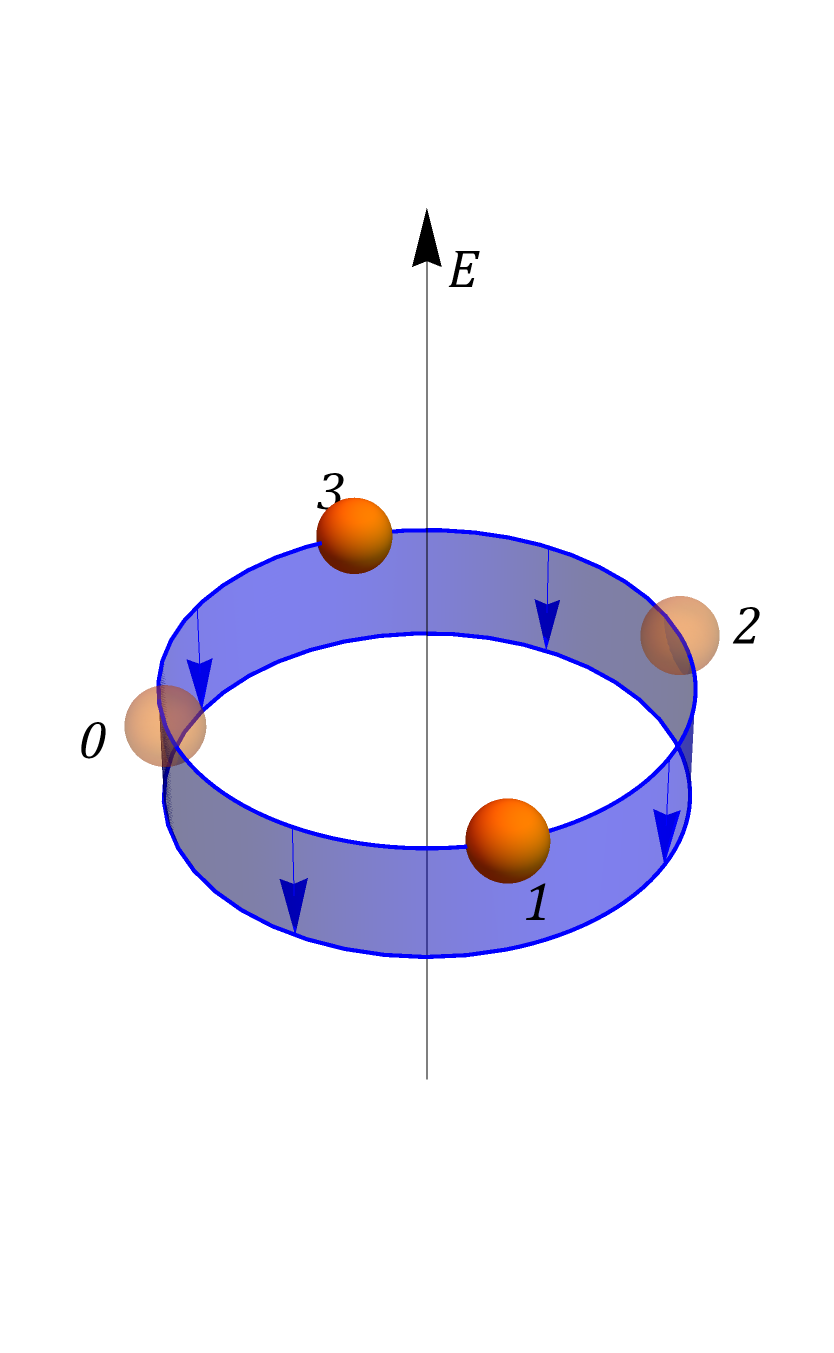}};
\draw[->,thick] (russell.20) to [bend right=-60] (whitehead.160)
    node [] (t1) at (1.6,1) {$\small{\mathcal{C}_+}$};
\draw[<-,thick] (russell.345) to [bend right=60] (whitehead.195)
    node [] (t1) at (1.6,-1) {$\small{\mathcal{C}_-}$};
\end{tikzpicture}}
    \end{tabular} \\
    \vspace{0.1cm}
    \begin{tabular}{c}
        \subfigure[]{\begin{tikzpicture}
\node[inner sep=0pt] (russell) at (0,0)
    {\includegraphics[width=.12\linewidth]{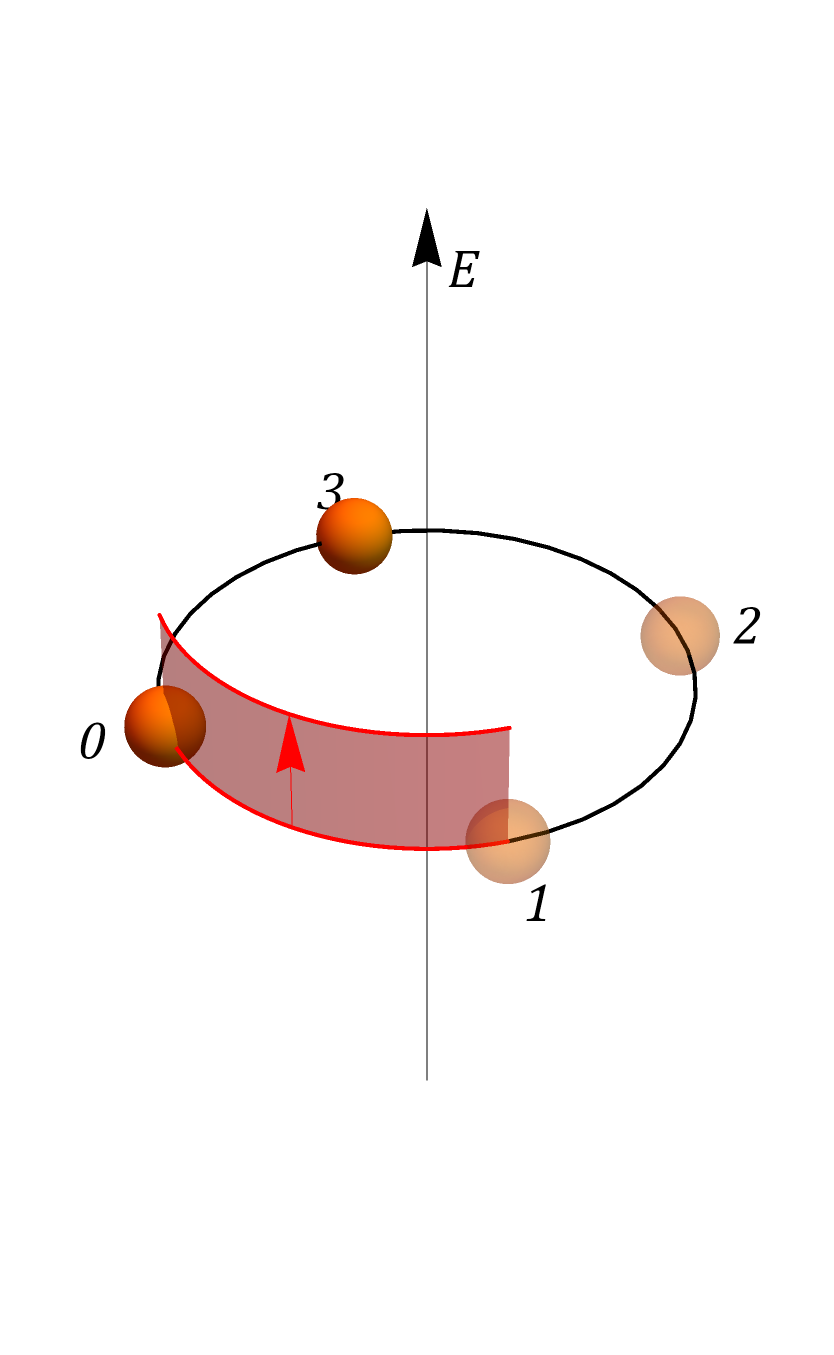}};
\node[inner sep=0pt] (whitehead) at (3,0) {\includegraphics[width=.12\linewidth]{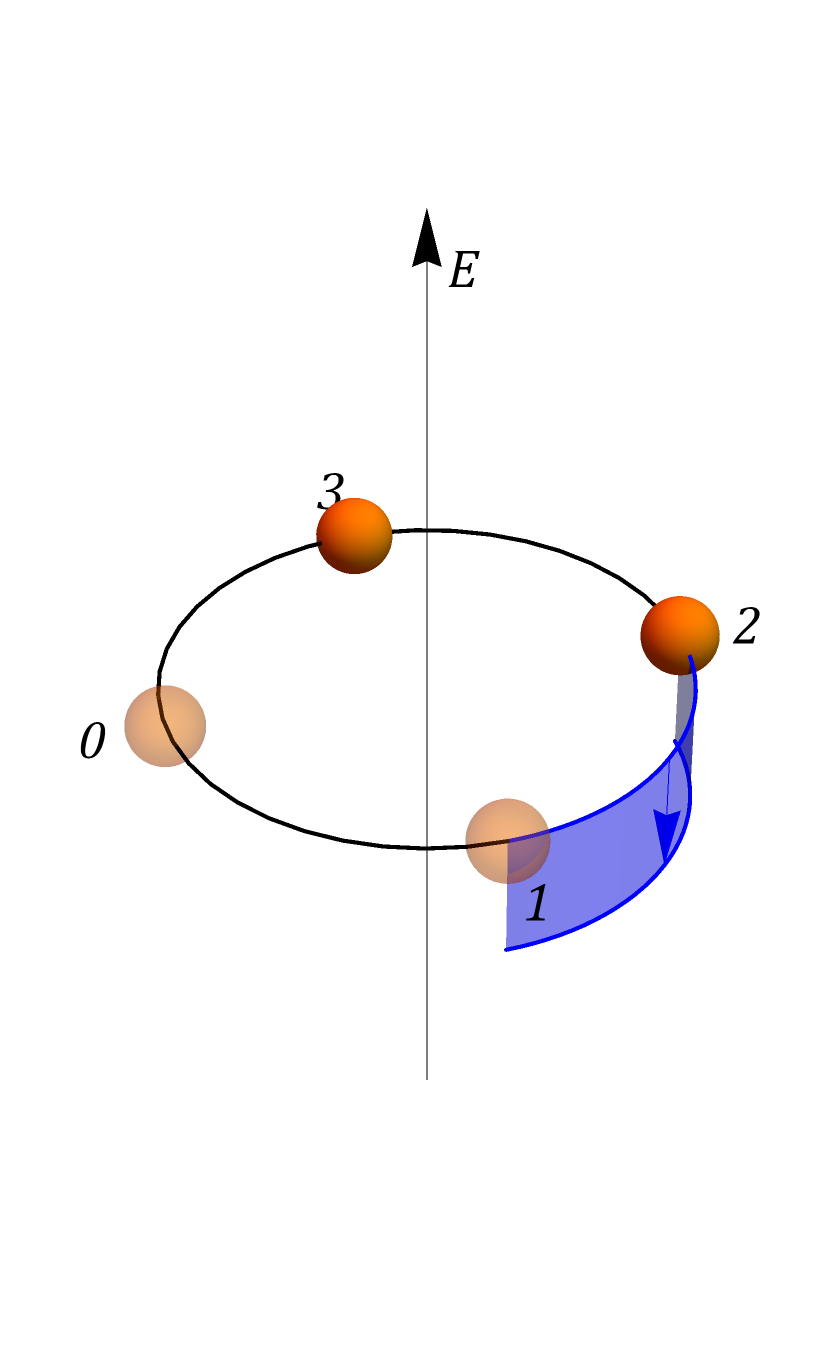}};
\node[inner sep=0pt] (russell2) at (6,0)
    {\includegraphics[width=.12\linewidth]{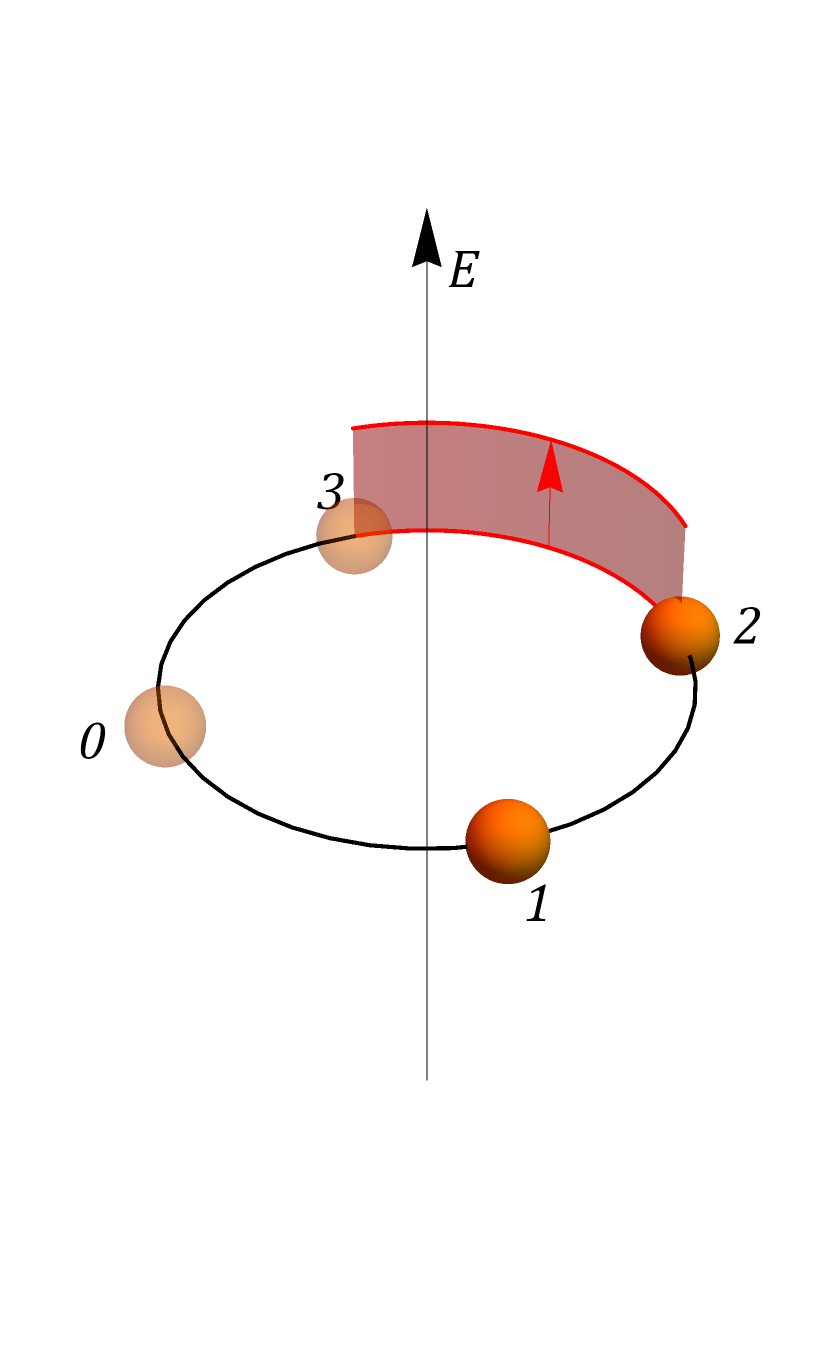}};
\node[inner sep=0pt] (whitehead2) at (9,0) {\includegraphics[width=.12\linewidth]{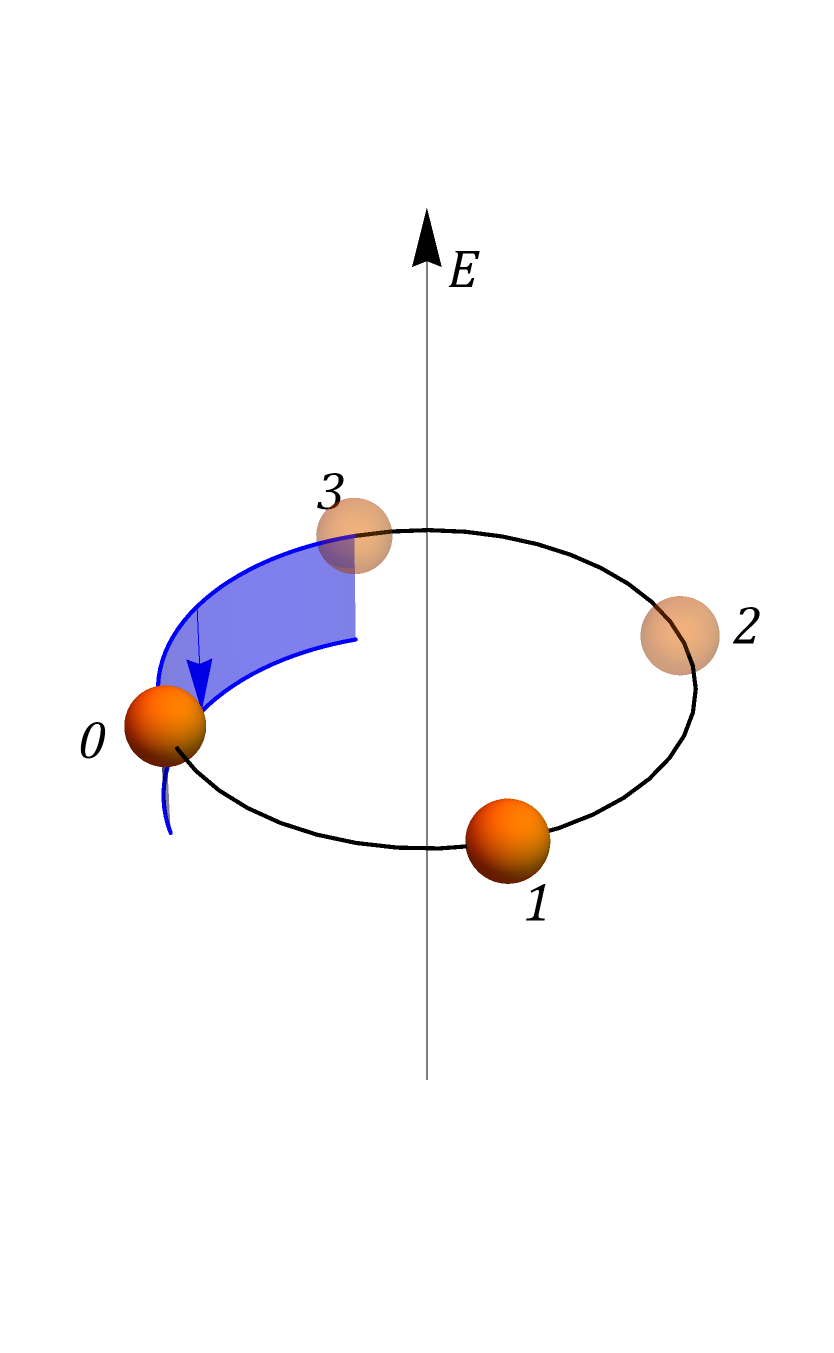}};
\draw[->,thick] (russell.20) to [bend right=-60] (whitehead.160)
    node [] (t1) at (1.6,1) {$\small{\mathcal{C}_+}$};
\draw[<-,thick] (russell.345) to [bend right=60] (whitehead.195)
    node [] (t1) at (1.6,-1) {$\small{\mathcal{C}_-}$};
\draw[->,thick] (whitehead.20) to [bend right=-60] (russell2.160)
    node [] (t1) at (4.6,1) {$\small{\mathcal{C}_+}$};
\draw[<-,thick] (whitehead.345) to [bend right=60] (russell2.195)
    node [] (t1) at (4.6,-1) {$\small{\mathcal{C}_-}$};
\draw[->,thick] (russell2.20) to [bend right=-60] (whitehead2.160)
    node [] (t1) at (7.6,1) {$\small{\mathcal{C}_+}$};
\draw[<-,thick] (russell2.345) to [bend right=60] (whitehead2.195)
    node [] (t1) at (7.6,-1) {$\small{\mathcal{C}_-}$};
\end{tikzpicture}}  
    \end{tabular} \\
    \vspace{0.1cm}
    \begin{tabular}{c}
        \subfigure[]{\begin{tikzpicture}
\node[inner sep=0pt] (russell) at (0,0)
    {\includegraphics[width=.12\linewidth]{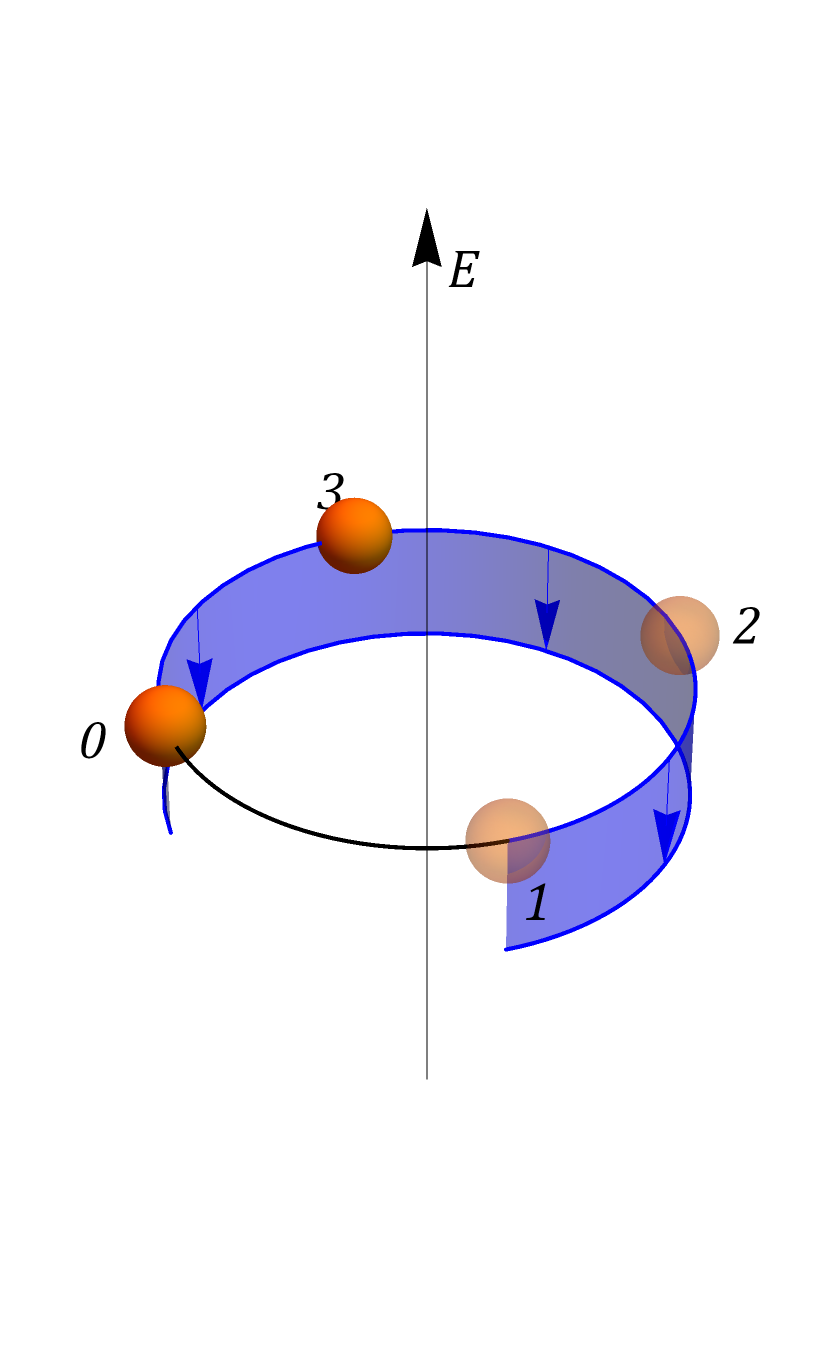}};
\node[inner sep=0pt] (whitehead) at (3,0) {\includegraphics[width=.12\linewidth]{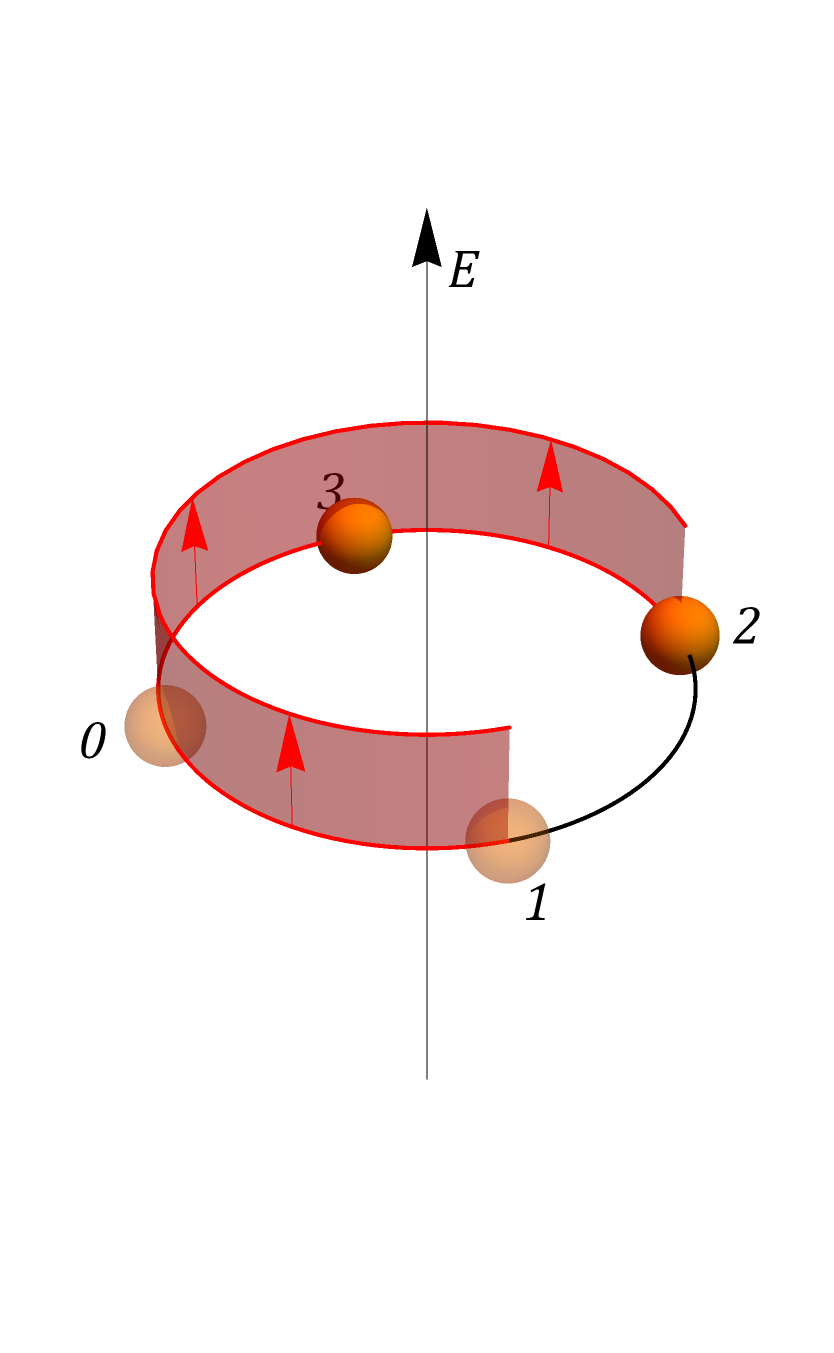}};
\node[inner sep=0pt] (russell2) at (6,0)
    {\includegraphics[width=.12\linewidth]{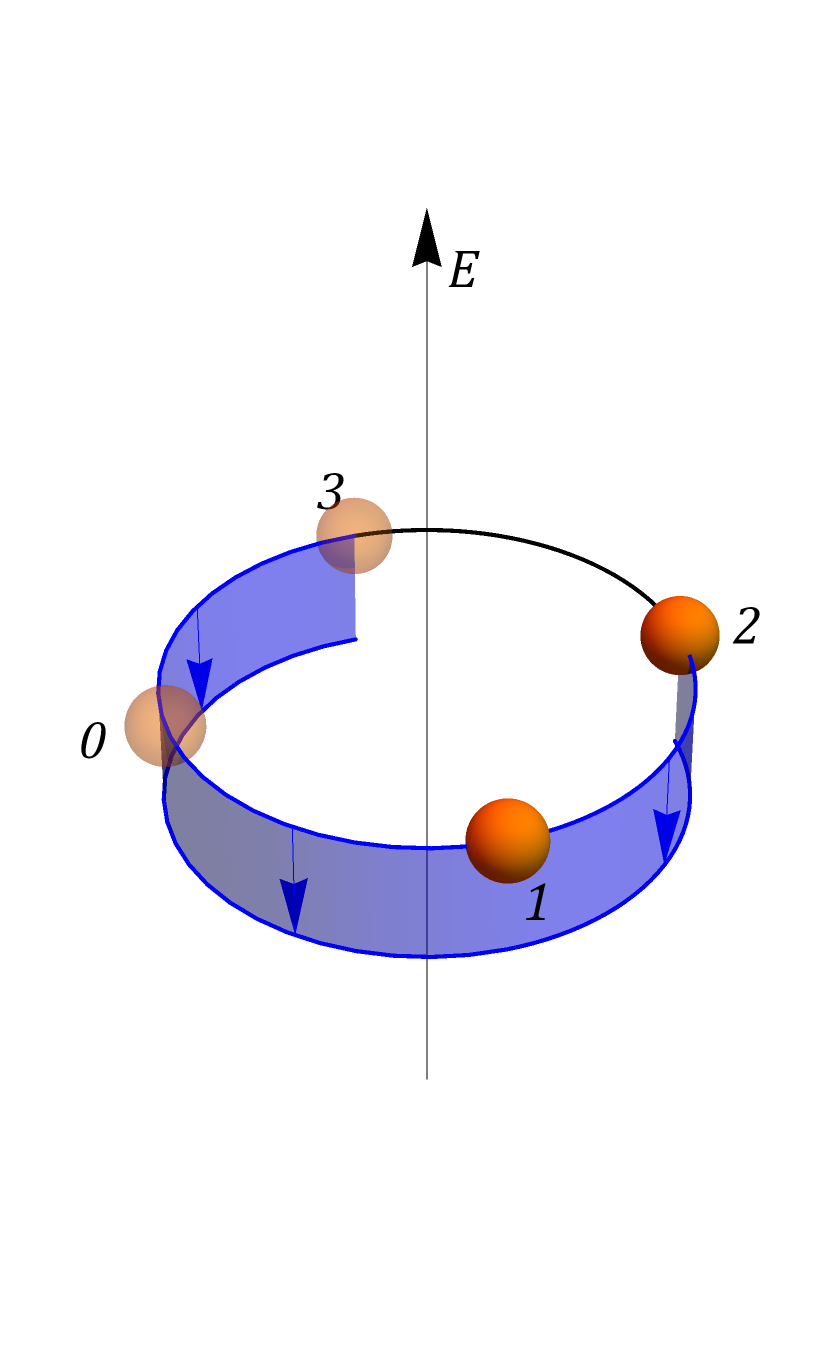}};
\node[inner sep=0pt] (whitehead2) at (9,0) {\includegraphics[width=.12\linewidth]{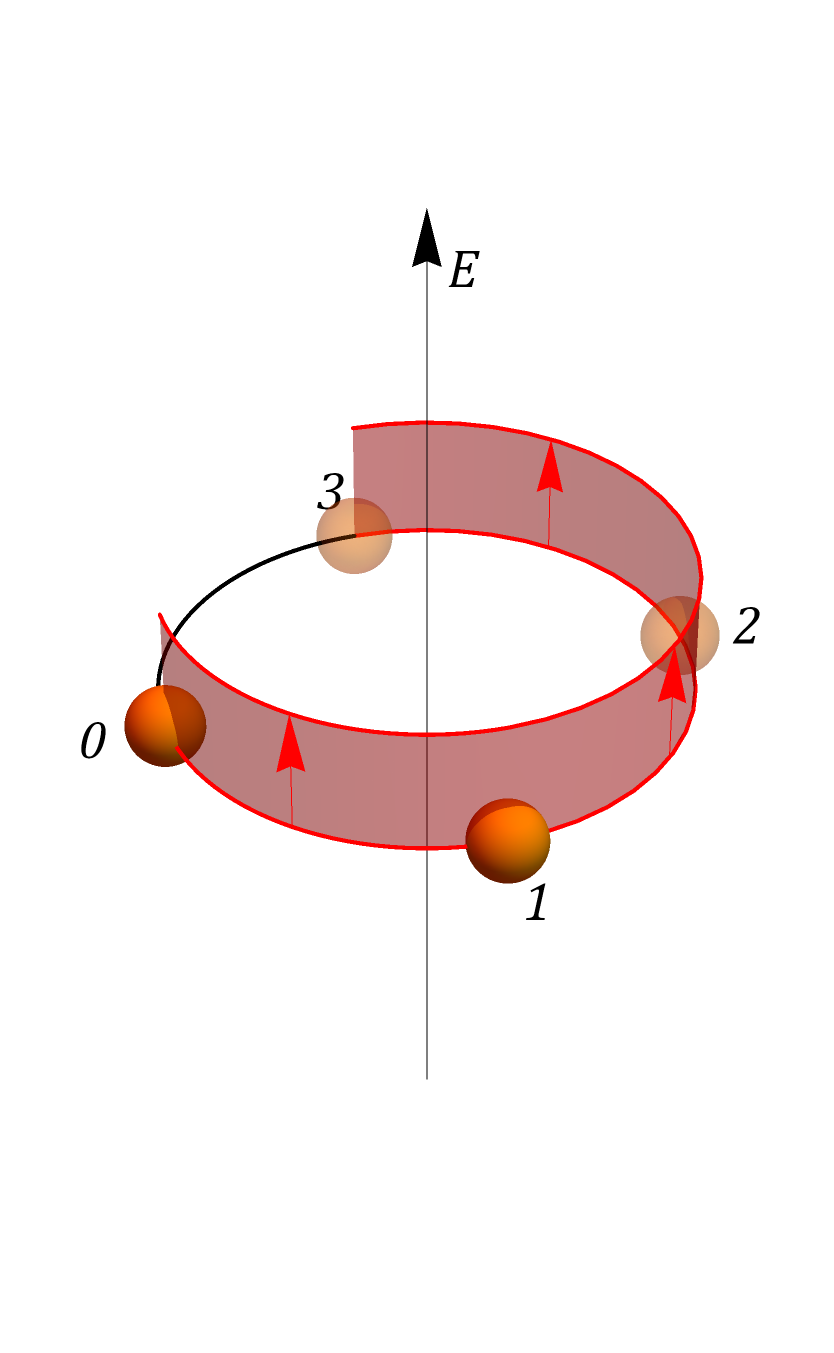}};
\draw[->,thick] (russell.20) to [bend right=-60] (whitehead.160)
    node [] (t1) at (1.6,1) {$\small{\mathcal{C}_+}$};
\draw[<-,thick] (russell.345) to [bend right=60] (whitehead.195)
    node [] (t1) at (1.6,-1) {$\small{\mathcal{C}_-}$};
\draw[->,thick] (whitehead.20) to [bend right=-60] (russell2.160)
    node [] (t1) at (4.6,1) {$\small{\mathcal{C}_+}$};
\draw[<-,thick] (whitehead.345) to [bend right=60] (russell2.195)
    node [] (t1) at (4.6,-1) {$\small{\mathcal{C}_-}$};
\draw[->,thick] (russell2.20) to [bend right=-60] (whitehead2.160)
    node [] (t1) at (7.6,1) {$\small{\mathcal{C}_+}$};
\draw[<-,thick] (russell2.345) to [bend right=60] (whitehead2.195)
    node [] (t1) at (7.6,-1) {$\small{\mathcal{C}_-}$};
\end{tikzpicture}}  
    \end{tabular} \\
    \vspace{0.1cm}
    \begin{tabular}{c}
        \subfigure[]{\begin{tikzpicture}
\node[inner sep=0pt] (russell) at (0,0)
    {\includegraphics[width=.12\linewidth]{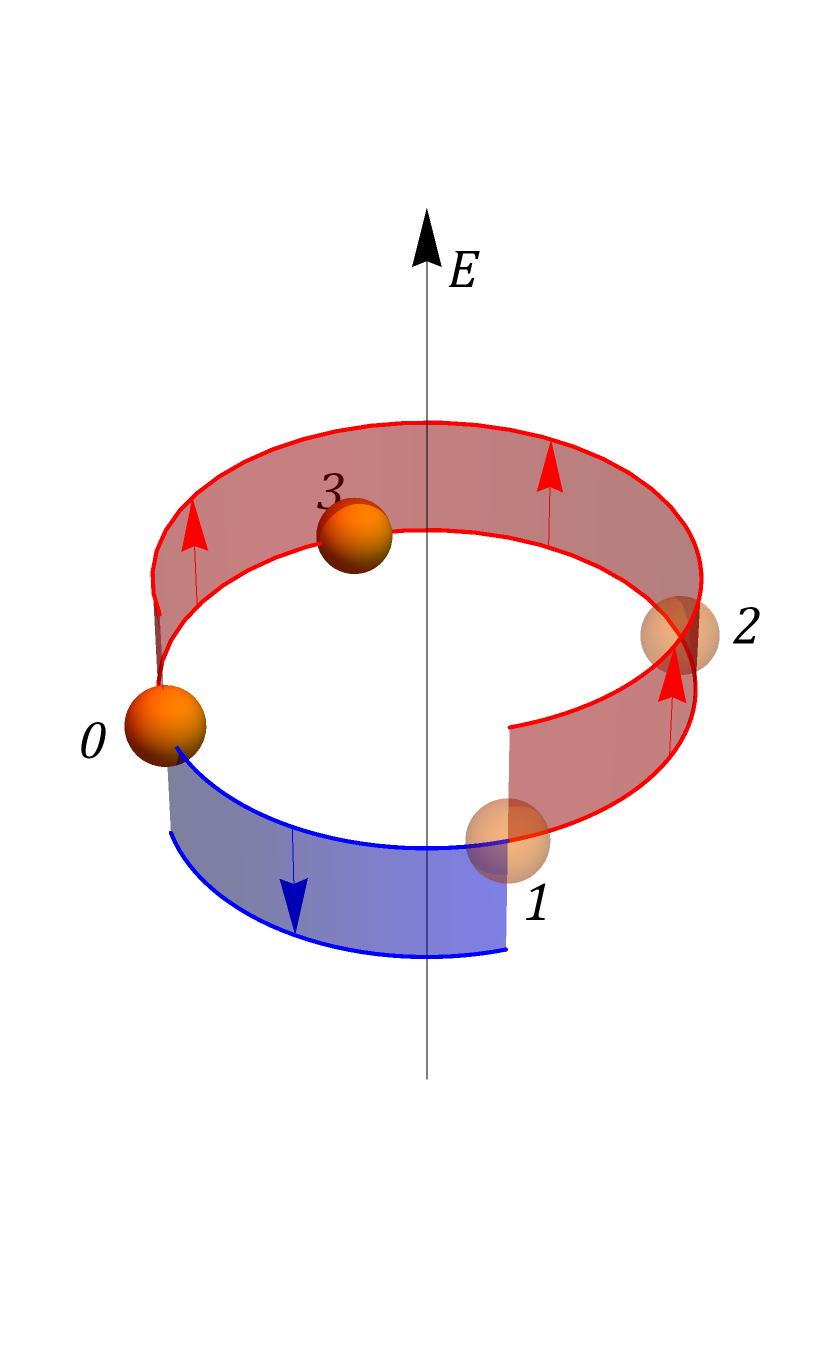}};
\node[inner sep=0pt] (whitehead) at (3,0) {\includegraphics[width=.12\linewidth]{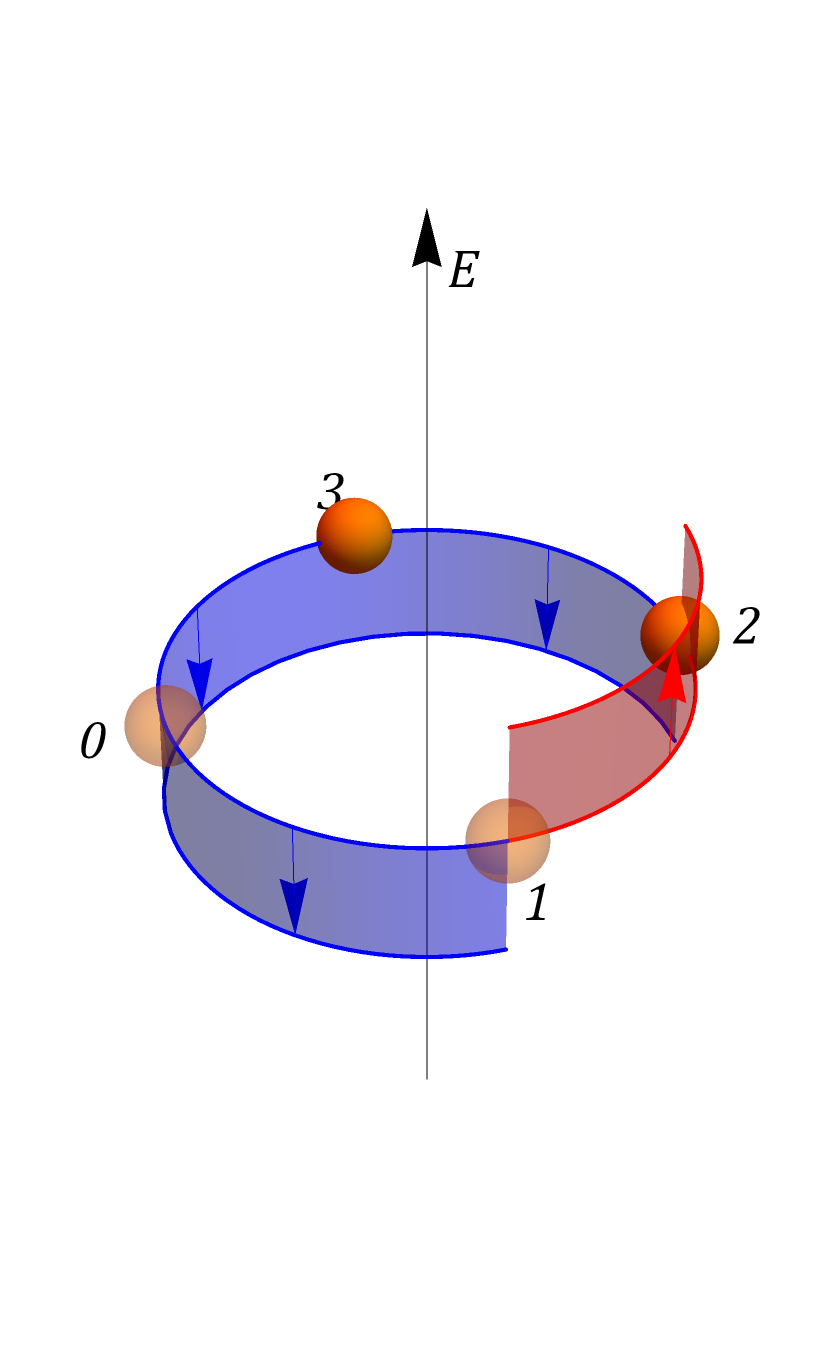}};
\node[inner sep=0pt] (russell2) at (6,0)
    {\includegraphics[width=.12\linewidth]{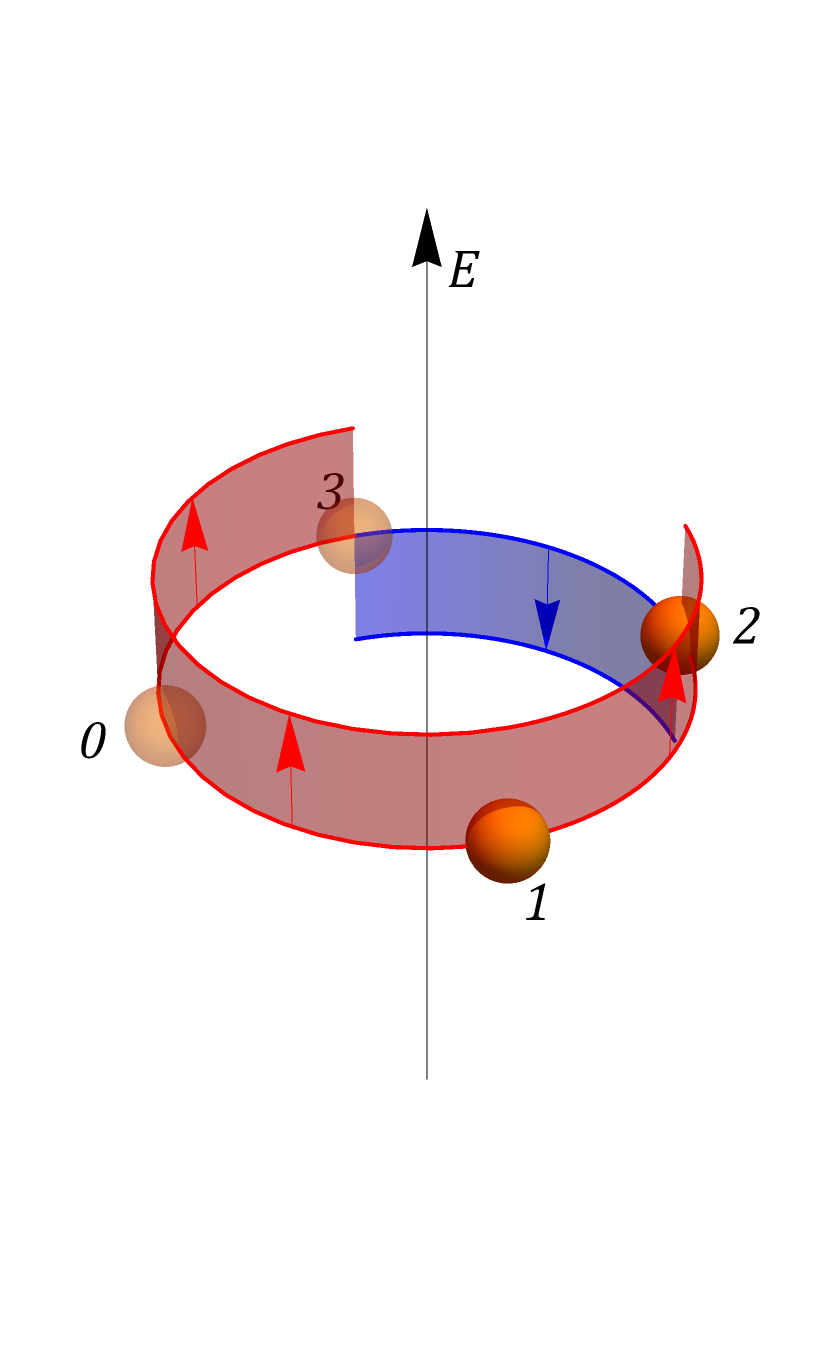}};
\node[inner sep=0pt] (whitehead2) at (9,0) {\includegraphics[width=.12\linewidth]{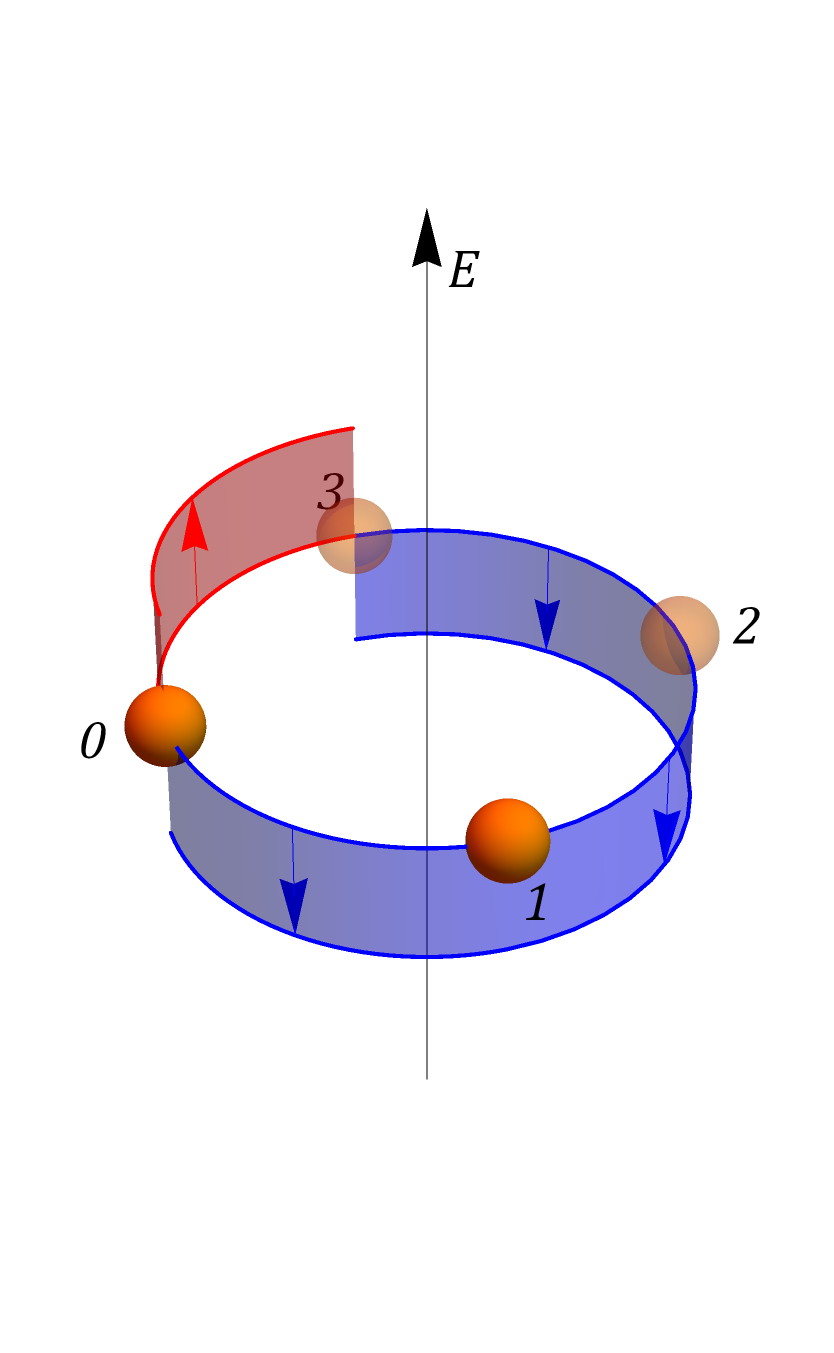}};
\draw[->,thick] (russell.20) to [bend right=-60] (whitehead.160)
    node [] (t1) at (1.6,1) {$\small{\mathcal{C}_+}$};
\draw[<-,thick] (russell.345) to [bend right=60] (whitehead.195)
    node [] (t1) at (1.6,-1) {$\small{\mathcal{C}_-}$};
\draw[->,thick] (whitehead.20) to [bend right=-60] (russell2.160)
    node [] (t1) at (4.6,1) {$\small{\mathcal{C}_+}$};
\draw[<-,thick] (whitehead.345) to [bend right=60] (russell2.195)
    node [] (t1) at (4.6,-1) {$\small{\mathcal{C}_-}$};
\draw[->,thick] (russell2.20) to [bend right=-60] (whitehead2.160)
    node [] (t1) at (7.6,1) {$\small{\mathcal{C}_+}$};
\draw[<-,thick] (russell2.345) to [bend right=60] (whitehead2.195)
    node [] (t1) at (7.6,-1) {$\small{\mathcal{C}_-}$};
\end{tikzpicture}}  
    \end{tabular}\\ 
    \vspace{0.1cm}
    \begin{tabular}{cc} 
        \subfigure[]{\includegraphics[width=.12\linewidth]{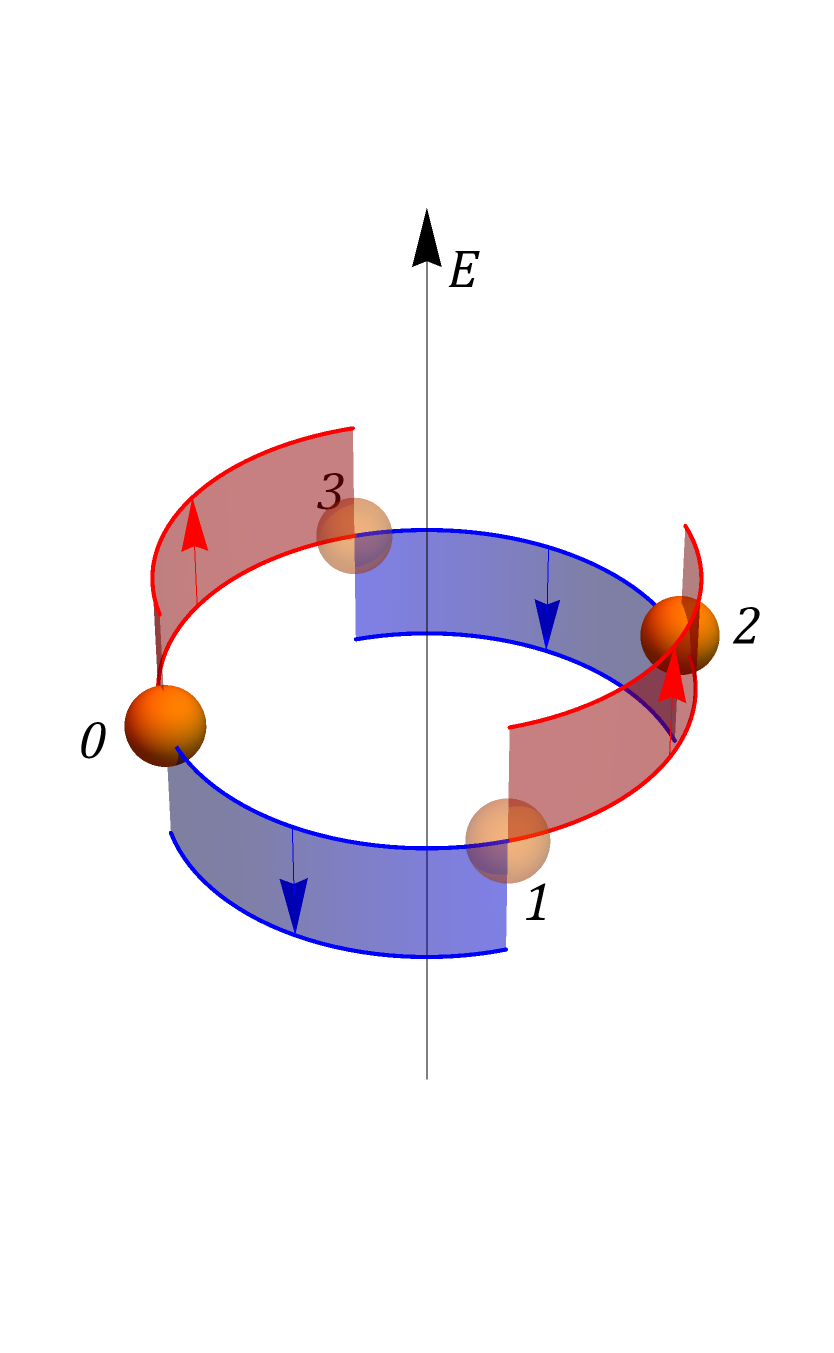}} & \subfigure[]{\begin{tikzpicture}
\node[inner sep=0pt] (russell) at (0,0)
    {\includegraphics[width=.12\linewidth]{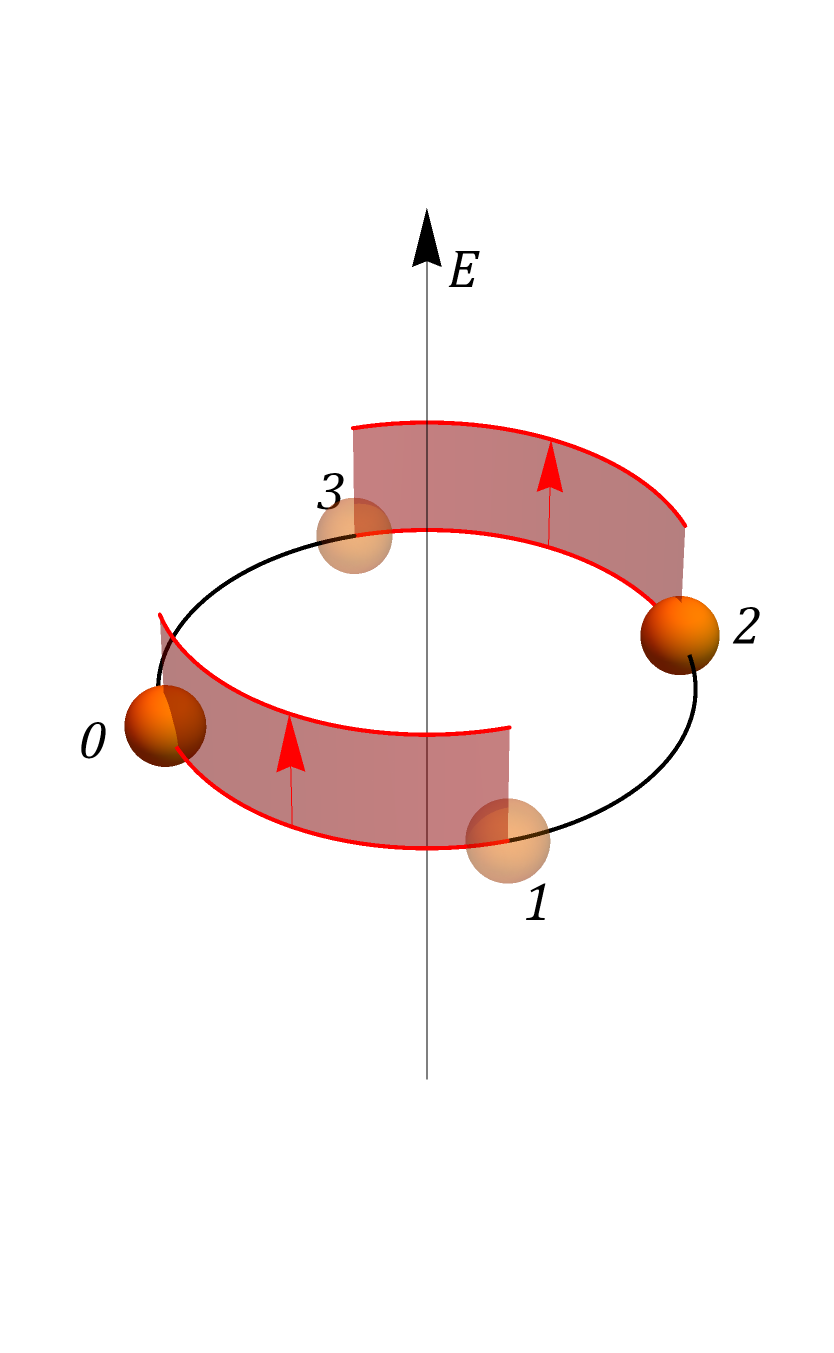}};
\node[inner sep=0pt] (whitehead) at (3,0) {\includegraphics[width=.12\linewidth]{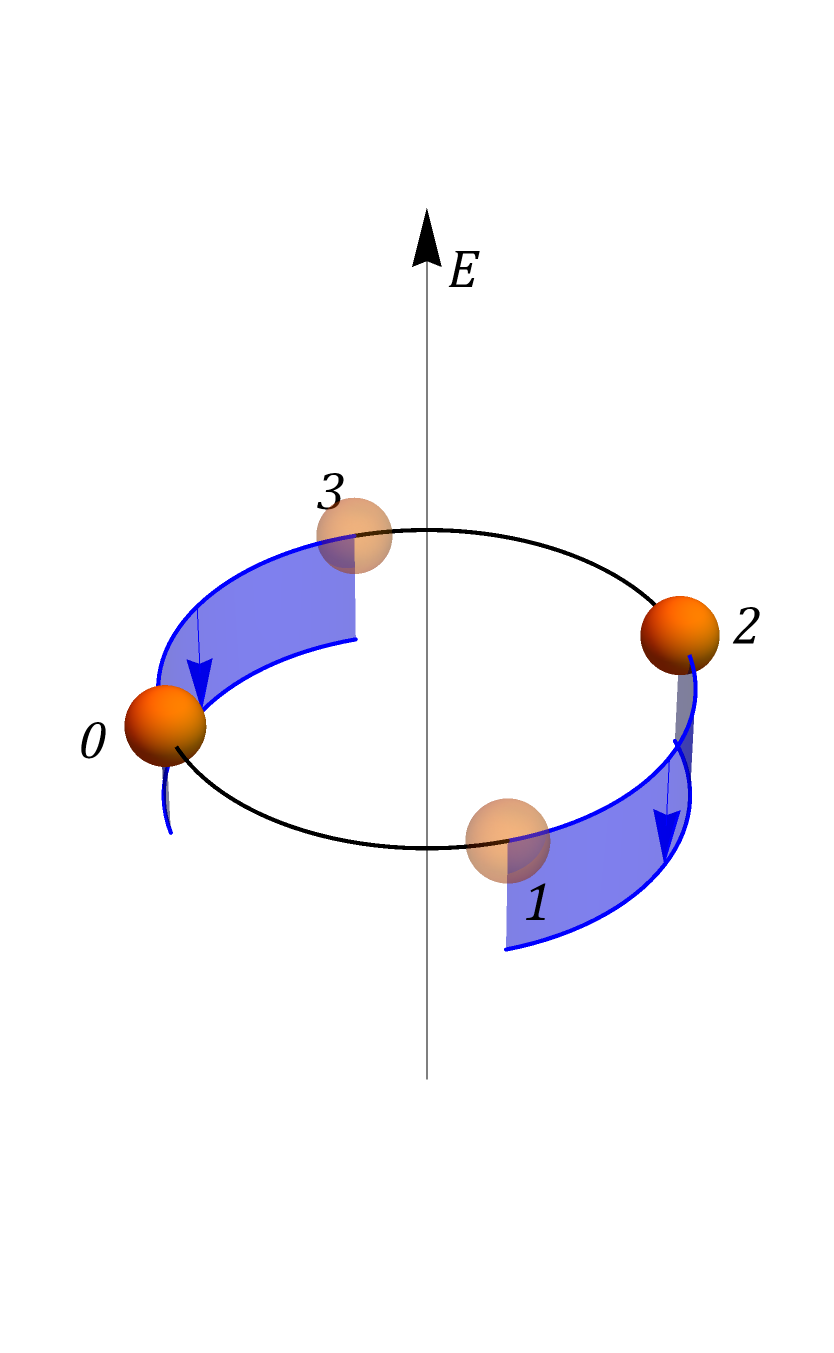}};
\draw[->,thick] (russell.20) to [bend right=-60] (whitehead.160)
    node [] (t1) at (1.6,1) {$\small{\mathcal{C}_+}$};
\draw[<-,thick] (russell.345) to [bend right=60] (whitehead.195)
    node [] (t1) at (1.6,-1) {$\small{\mathcal{C}_-}$};
\end{tikzpicture}} 
    \end{tabular} 
    \end{tabular}
    \caption{A basis of gauge invariant states for the periodic lattice with $N=4$. Panels (a,b) represents the Dirac vacuum states; panels (c,d,e) one-meson states and panels (f,g) two-meson states. Action of the charge conjugation operators $\mathcal{C}_{\pm}$ and the corresponding orbits are shown.}
    \label{fig:orbits}
\end{figure}

All the 18 states are shown in Fig.~\ref{fig:orbits}, where:
\begin{enumerate}
\item panels (a,b) show the Dirac vacuum, corresponding to the two even sites empty and the two odd sites being filled,  with the three possible total background fields, which are completely fixed by the value of the electric field $e_0$ on the edge joining the sites $0$ and $1$: the state with zero electric field background is a fixed point with respect to ${\cal C}_+$, while the other two states with nonzero electric field background give one orbit. \\
\item panels (c,d,e) show the 12 one-meson states, corresponding to occupying one even and one odd site, the three panels corresponding to the orbits of the three possible configurations obtained from the three possible Dirac vacuum configurations, respectively, by moving a particle  from site $2$ to site $3$;\\
\item panels (f,g) show the two-meson states, corresponding to the two even sites being filled and the two odd sites being empty  and the three possible total field configurations, the two states with $|e_0 \rangle = |0\rangle, |+\rangle$ represent an orbit, while the state with $|e_0 \rangle = |-\rangle$ is a fixed point. 
\end{enumerate}

In order to achieve an optimal encoding, we exploit the translation and charge conjugation symmetry operators that commute with the Hamiltonian, $\mathcal{T}_2 \mathcal{H} \mathcal{T}_2 = \mathcal{H}$ and $\mathcal{C}_{\pm} \mathcal{H} \mathcal{C}_{\mp} = \mathcal{H}$, to define subspace on which the dynamics is restricted. The elements of a basis for such subspaces are obtained as linear combinations of the states shown in Fig.~\ref{fig:orbits} as follows:\\
\begin{itemize}
\item subspace with $T_2 = +1; ~ C_{+} = +1$
\begin{align}
&\ket{\psi_1} = \ket{0000} \label{A1}\\
&\ket{\psi_2} =\frac{1}{2}(\ket{+000}+\ket{000-}+\ket{0-00}+\ket{00+0}) \\
&\ket{\psi_3} =\frac{1}{\sqrt{2}}(\ket{+0+0} + \ket{0-0-}) \\
&\ket{\psi_4} =\frac{1}{2}(\ket{+0++}+\ket{+++0}+\ket{--0-}+\ket{0---}) \\
&\ket{\psi_5} =\frac{1}{\sqrt{2}}(\ket{++++} + \ket{----}) \\
&\ket{\psi_6} =\frac{1}{2}(\ket{-+++}+\ket{++-+}+\ket{---+}+\ket{-+--}) \\
&\ket{\psi_7} =\ket{-+-+} \label{A7}
\end{align}
\item subspace with $T_2 = +1; ~ C_{+} = -1$
\begin{align}
&\ket{\psi_8} =\frac{1}{2}(\ket{+000}-\ket{000-}-\ket{0-00}+\ket{00+0}) \\
&\ket{\psi_9} =\frac{1}{\sqrt{2}}(\ket{+0+0} - \ket{0-0-}) \\
&\ket{\psi_{10}} =\frac{1}{2}(\ket{+0++}+\ket{+++0}-\ket{--0-}-\ket{0---}) \\
&\ket{\psi_{11}} =\frac{1}{\sqrt{2}}(\ket{++++} - \ket{----}) \\
&\ket{\psi_{12}} =\frac{1}{2}(\ket{-+++}+\ket{++-+}-\ket{---+}-\ket{-+--})
\end{align}
\item subspace with $T_2 = -1; ~ C_{+} = +\ii$
\begin{align}
&\ket{\psi_{13}} =\frac{1}{2}(\ket{+000}+\ii\ket{000-}-\ii\ket{0-00}-\ket{00+0}) \\
&\ket{\psi_{14}} =\frac{1}{2}(\ket{+0++}-\ket{+++0}-\ii\ket{--0-}+\ii\ket{0---}) \\
&\ket{\psi_{15}} =\frac{1}{2}(\ket{-+++}-\ket{++-+}+\ii\ket{---+}-\ii\ket{-+--})
\end{align}
\item subspace with $T_2 = -1; ~ C_{+} = -\ii$
\begin{align}
&\ket{\psi_{16}} =\frac{1}{2}(\ket{+000}-\ii\ket{000-}+\ii\ket{0-00}-\ket{00+0}) \\
&\ket{\psi_{17}} =\frac{1}{2}(\ket{+0++}-\ket{+++0}+\ii\ket{--0-}-\ii\ket{0---}) \\
&\ket{\psi_{18}} =\frac{1}{2}(\ket{-+++}-\ket{++-+}-\ii\ket{---+}+\ii\ket{-+--})
\end{align}
\end{itemize}
The Hamiltonian diagonal block presented in Eq. \eqref{Eq::H++} describes the dynamics in the sector spanned by eigenvectors in the subspace with $T_2 = +1; ~ C_{+} = +1$. The remaining ones in the diagonal blocks structure $U \mathcal{H} U^\dagger = \mathcal{H}^{(+,+)} \oplus \mathcal{H}^{(+,-)}\oplus \mathcal{H}^{(-,+\ii)}\oplus \mathcal{H}^{(-,-\ii)}$ are not involved in the time evolution of the initial state that we fix to be the Dirac vacuum, but, for completeness, we show all of them:
\begin{equation}
    \mathcal{H}^{(+,-)} = \begin{pmatrix}
        \frac{\pi}{3} & \frac{\xi}{\sqrt{2}} & & & \\
        \frac{\xi}{\sqrt{2}} & 2\mu + \frac{2\pi}{3} & \frac{\xi}{\sqrt{2}} & & \\
         & \frac{\xi}{\sqrt{2}} & \pi & \frac{\xi}{\sqrt{2}} & \\
         & & \frac{\xi}{\sqrt{2}} & -2\mu + \frac{4\pi}{3} & \frac{\xi}{\sqrt{2}} \\
         & & & \frac{\xi}{\sqrt{2}} & \frac{4\pi}{3} \\
    \end{pmatrix}, \ \
    \mathcal{H}^{(-,+\ii)} = \mathcal{H}^{(-,-\ii)} = \mathrm{diag} \left\{\frac{\pi}{3},\pi,\frac{4\pi}{3}\right\}.
\end{equation}

\section[\appendixname~\thesection]{Optimal permutation \label{secB1}}

With the simplest embedding, the Hamiltonian in the $(+,+)$ sector is written according to formula \eqref{Eq::Hnoperm}, where the following are the 20 non-zero coefficients:
\begin{center}
\begin{tabular}{ccccc}
    $c_{(0,0,0)} = \frac{3\pi}{4}$ & $c_{(0,0,1)} = \frac{(1+\sqrt{2})\xi}{4}$ & $c_{(0,0,3)} = \frac{\pi}{12}$ & $c_{(0,1,1)} = \frac{(2+\sqrt{2})\xi}{8}$ & $c_{(1,1,1)} = \frac{\xi}{4\sqrt{2}}$ \\
    $c_{(0,3,0)} = -\mu$ & $c_{(0,3,3)} = -\frac{\pi}{6}-\mu$ & $c_{(0,3,1)} = \frac{\xi}{4}$ & $c_{(0,2,2)} = \frac{(2+\sqrt{2})\xi}{8}$ & $c_{(1,2,2)} = -\frac{\xi}{4\sqrt{2}}$ \\
    $c_{(3,0,0)} = -\frac{\pi}{4}$ & $c_{(3,0,3)} = -\frac{\pi}{4}$ & $c_{(3,0,1)} = \frac{\xi}{4}$ & $c_{(3,1,1)} = \frac{(\sqrt{2}-2)\xi}{8}$ & $c_{(2,1,2)} = \frac{\xi}{4\sqrt{2}}$ \\
    $c_{(3,3,0)} = -\frac{\pi}{3}$ & $c_{(3,3,1)} = \frac{(\sqrt{2}-1)\xi}{4}$ & $c_{(3,3,3)} = \frac{\pi}{6}$ & $c_{(3,2,2)} = \frac{(\sqrt{2}-2)\xi}{8}$ & $c_{(2,2,1)} = \frac{\xi}{4\sqrt{2}}$ \\
\end{tabular}\\
\end{center}
The term with $c_{(0,0,0)}$ is just a phase and it can be safely neglected. Then there are 4 single-qubit, 8 two-qubit and 7 three-qubit terms.

\begin{figure}[t!]
\centering
\includegraphics[width=0.6\linewidth]{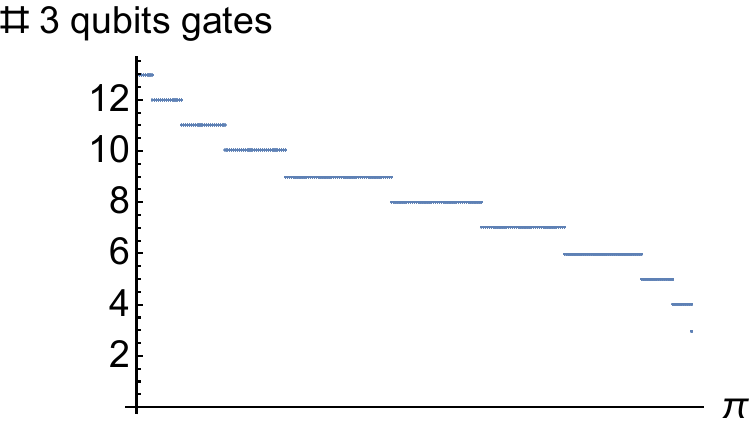}
\caption{Number of triple-qubits gates implied by the embedded Hamiltonian decomposition \eqref{Eq::Hnoperm} with respect to any permutation $\pi \in S_8$. Permutations are ordered with respect to the decreasing number of gates. }
\label{fig:num_3q}
\end{figure}

The form of the Hamiltonian can be considerably simplified if we modify the embedding by considering the optimal permutation of the indices $\pi_o = (7,6,1,2,4,5,8,3)$, which  is characterized by the same phase term and the following non-zero coefficients:
\begin{center}
\begin{tabular}{ccccc}
    $c_{(0,0,3)} = -\frac{\pi}{6}$ & $c_{(1,0,0)} = \frac{\xi}{2\sqrt{2}}$ & $c_{(1,0,3)} = -\frac{\xi}{2\sqrt{2}}$ & $c_{(0,3,3)} = \frac{\pi}{12}+\mu$ & $c_{(0,1,1)} = \frac{\xi}{4\sqrt{2}}$ \\
    $c_{(0,3,0)} = \frac{\pi}{2}$ & $c_{(3,3,0)} = \frac{\pi}{12}+\mu$ & $c_{(0,2,2)} = -\frac{\xi}{4\sqrt{2}}$ & $c_{(0,0,1)} = \frac{(4+\sqrt{2})\xi}{8}$ & $c_{(0,3,1)} = \frac{\xi}{4\sqrt{2}}$ \\
    $c_{(3,0,3)} = \frac{\pi}{12}$ & $c_{(3,3,1)} = -\frac{\xi}{4\sqrt{2}}$ & $c_{(3,1,1)} = -\frac{\xi}{4\sqrt{2}}$ & $c_{(3,0,1)} = \frac{(4-\sqrt{2})\xi}{8}$ & $c_{(3,2,2)} = \frac{\xi}{4\sqrt{2}}$ \\
\end{tabular}\\
\end{center}
yielding 4 single-qubit, 8 two-qubit and 3 three-qubit gates. For the problem under consideration, such optimal permutation can be simply obtained by testing all possible $8!$ permutations, as shown in Fig. \ref{fig:num_3q} which gives the number of terms with a product of three-qubit gates, with permutations ordered from the maximum number to the lowest, which is 3 in this case. 

With this permutation, the particle density operator is given by Eq. (\ref{Eq::num}) with 8 non-zero coefficients given by
$d_{(0,0,0)} = \frac{7}{16}, d_{(0,3,3)} = \frac{5}{16}, d_{(3,3,0)} = \frac{3}{16}, d_{(0,3,0)} = d_{(3,0,0)} = d_{(3,0,3)} = \frac{1}{16}, d_{(0,0,3)} = d_{(3,3,3)} = -\frac{1}{16}$. This allows to write:
\begin{equation} \tag{B1}
   \widetilde{\nu}^{(+,+)} = d_{(0,0,0)} \mathbb I + \sum_{\alpha=1}^7  P_\alpha      \label{nuespl}
\end{equation}   
 where the seven operators $P_\alpha$ are given by the seven non-zero coefficients $d_{(i,j,k)}$  multiplied by the corresponding triple-string of Pauli operators  $\sigma_i \otimes \sigma_j \otimes \sigma_k$ not equal to the identity.

 \section[\appendixname~\thesection]{General procedure for classical-quantum embedding \label{secBB1}}
 
 For future research, it is important to design an algorithm to find the gauge invariant subspace and obtain the optimal permutation that is able to perform fast also when we scale up the lattice size $N$.  The basic procedure is as follows:
\begin{itemize}
\item Step 1:  Store a list of possible site configurations for all even and odd sites. This is a total of 6 configurations each for both even and odd sites since:\\
- each site, represented as $x$, can be filled or empty; \\
- for either value of occupancy, the link to the right, represented as $e_{x,x+1}$, can take three possible values of electric field. The link to the left, $e_{x-1,x}$, is fixed by Gauss' law.

\item Step 2: Place sites ${0,...N-1}$ one after the other, following the pattern even-odd-even-odd... ensuring that the values of the common links match.

\item Step 3: Impose periodic boundary conditions: $e_{N,N+1} = e_{-1,0}$.

\item Step 4: Eliminate the configurations that do not correspond to the right total number of particles (in our case, half filling). After this step,  we obtain basis states which are best represented as a dictionary, with each site mapping to its occupancy ($0$ or $1$), and each link mapping to a $\mathbb{Z}_3$ element representing the electric field on that link (with values $-1,0,+1$).

\item Step 5: Arrange states into multiplets under translation and charge conjugation symmetries and isolate the multiplet containing the vacuum state. This multiplet corresponds the $(+,+)$ sector and evolves independently of the other blocks.

\item Step 6: Find the Hamiltonian ${\cal H}\equiv {\cal H}^{(+,+)}$ matrix elements in this multiplet. Embed this matrix into a $2^M$-dimensional space where $M$ is the lowest such value the matrix can be embedded into. This gives the number of needed qubits.

\item Step 7: Find an optimal permutation of the $2^M$ basis states such that the number of higher-qubit gates in the Pauli gate expansion of ${\cal H}$ is minimized.

\item Step 8: Use the Pauli gate expansion of the Hamiltonian in the optimized encoding to construct the Trotter steps in the time evolution. 
\end{itemize}

For large system sizes, it is unfeasible to perform Step 7 using a brute-force approach like we have done for the case of $N=4$. This is a discrete optimization problem where we have to find the optimal permutation of basis states. The entire search space is $M!$ which grows faster than exponential, and it quickly becomes unfeasible to perform an exhaustive search.\\
The approach we take is a local search algorithm, that can be outlined as follows:
\begin{itemize}
	\item We define the objective function as
\begin{equation} \tag{C1}
    \mathcal{F}(H)= \sum_i Q[\text{Tr}({\cal H} X_i)], \quad Q(x)\equiv
\begin{cases}
    1,& \text{if } x\neq 0\\
    0,              & \text{otherwise}
\end{cases} ,
\end{equation}
where $\{X_i\}$ is the set of $M$-qubit Pauli gates. This simply counts the number of such gates in the decomposition of the Hamiltonian ${\cal H}$. 
    \item Start with an initial configuration of the Hamiltonian ${\cal H}_{init}$ and store $f_{opt}=\mathcal{F}({\cal H}_{init})$.
    \item We choose a ``neighbourhood" of a configuration of ${\cal H}$ to be the set of configurations related to it by a single swap of two indices: row $i \longleftrightarrow$ row $j$, column $i \longleftrightarrow$ column $j$. For a $d=2^M$-dimensional system, the dimension of each neighbourhood  scales as  $\mathcal{O}(d^2)$.
    \item Go to a locally optimal configuration in this neighbourhood, according to some heuristic choice; calculate the objective function for the new configuration and update $f_{best}$.
    \item Continue until $f_{best}$ cannot be improved any further.
\end{itemize}

We have tested this procedure with a few different lattice sizes $N=2,4,6,8$. For these cases, we have simply used a greedy local search algorithm \cite{permutation_gradient}. The optimal number of $M$-qubit gates so found and the runtimes are tabulated below:\\

\begin{center}
Table 1: Results of typical runs of the greedy algorithm
\begin{tabular}{||c c c c||} 
 \hline
 $L$ & System size $M$ & $M$-qubit gates in optimal permutation & Runtime \\ [0.5ex] 
 \hline\hline
 2 & 4 & 2 & $\sim 10^{-3}$ s \\ 
 \hline
 4 & 8 & 3 & $\sim 10^{-2}$ s \\
 \hline
 6 & 16 & 8 & $\sim 10^{-1}$ s \\
 \hline
 8 & 64 & 123 & $\sim 10^3$ s \\
 \hline
\end{tabular}
\end{center}
\vskip5mm
This local search algorithm gives us a ``good enough" solution, but there is no way of knowing if it is the true optimal solution. We can improve our results by running the algorithm a few times from random initial configurations and taking the best result.

 \section[\appendixname~\thesection]{Digital circuit for Trotter evolution \label{secB2}}
 
 We describe here the Trotter expansion of the evolution operator $U(t)$ corresponding to a time step $\Delta t$ of the optimal permuted Hamiltonian ${\cal H}$, framed in $\texttt{qiskit}$ notation. The decomposition in simpler operators proceeds as follows.
\begin{itemize}
\item Single-qubit gates are related to the coefficients with $c_{(0,0,1)}, c_{(0,0,3)}, c_{(0,3,0)}, c_{(1,0,0)}$. The are can implemented by the circuit
\[ \Qcircuit @C=1em @R=.7em {
	\lstick{q_0} & \qw & \qw & \qw & \gate{R_X\left(\frac{\xi \Delta t}{\sqrt{2}}\right)} & \qw \\
	\lstick{q_1}  & \qw & \qw & \gate{R_Z\left(\pi \Delta t\right)} & \qw & \qw \\
	\lstick{q_2}  & \gate{R_X\left(\frac{4+\sqrt{2}}{4} \xi \Delta t\right)} & \gate{R_Z\left(-\frac{\pi}{3} \Delta t\right)} & \qw & \qw & \qw
} \]
which is endowed with a high average fidelity. 
\item Two-qubit gates can be grouped in three different operators. The first one contains the terms corresponding to the coefficients $c_{(0,1,1)}, c_{(0,2,2)}, c_{(0,3,3)}$ and is represented by the circuit
\[ \Qcircuit @C=1em @R=.7em {
	\lstick{q_0} & \qw & \qw & \qw & \qw & \qw & \qw & \qw & \qw \\
    \lstick{q_1} & \ctrl{1} & \gate{R_X\left(\frac{\xi \Delta t}{2\sqrt{2}}\right)} & \ctrl{1} & \gate{R_X\left(\frac{\xi \Delta t}{2\sqrt{2}}\right)} & \ctrl{1} & \qw & \ctrl{1} & \qw \\
    \lstick{q_2} & \targ & \qw & \ctrl{-1} & \qw & \ctrl{-1} & \gate{R_Z\left((\frac{\pi}{6}+2\mu) \Delta t\right)} & \targ & \qw
}
\]
The second operator contains the terms corresponding to the coefficients $c_{(0,3,1)}, c_{(1,0,3)}$ and is given by the circuit
\[ \Qcircuit @C=1em @R=.7em {
	\lstick{q_0} & \qw & \qw & \qw & \gate{H} & \ctrl{2} & \qw & \ctrl{2} & \gate{H} & \qw \\
	\lstick{q_1}  & \ctrl{1} & \qw & \ctrl{1} & \qw & \qw & \qw & \qw & \qw & \qw \\
	\lstick{q_2}  & \gate{Y} & \gate{R_X\left(\frac{\xi \Delta t}{2\sqrt{2}}\right)} & \gate{Y} & \gate{H} & \gate{Y} & \gate{R_X\left(-\frac{\xi \Delta t}{\sqrt{2}}\right)} & \gate{Y} & \gate{H} & \qw
}
\]
Finally the third operator, which contains the terms with the coefficient $c_{(3,0,1)}, c_{(3,0,3)}, c_{(3,3,0)}$, corresponds to the circuit
\[ \Qcircuit @C=.7em @R=.7em {
	\lstick{q_0} & \ctrl{2} & \qw & \ctrl{2} & \ctrl{2} & \qw & \ctrl{2} & \ctrl{1} & \qw & \ctrl{1} & \qw \\
	\lstick{q_1}  & \qw & \qw & \qw & \qw & \qw & \qw & \targ & \gate{R_Z\left((\frac{\pi}{6}+2\mu) \Delta t\right)} & \targ & \qw \\
	\lstick{q_2}  & \gate{Y} & \gate{R_X\left(\frac{(4-\sqrt{2})\xi \Delta t}{4} \right)} & \gate{Y} & \targ & \gate{R_Z\left(\frac{\pi \Delta t}{6} \right)} & \targ & \qw & \qw & \qw & \qw
}
\]
Here, the abundance of controlled gates cause a decreasing average fidelity with respect to single-qubits gates case. 
\item Three-qubit gates are grouped in two different operators, The first one contains the terms with coefficients $c_{(3,3,1)}$ and is given by the circuit
\[ \Qcircuit @C=1em @R=.7em {
	\lstick{q_0} & \ctrl{1} & \qw & \qw & \qw & \ctrl{1} \\
	\lstick{q_1}  & \targ & \ctrl{1} & \qw & \ctrl{1} & \targ \\
	\lstick{q_2}  & \qw & \gate{Y} & \gate{R_X\left(-\frac{\xi \Delta t}{2\sqrt{2}}\right)} & \gate{Y} & \qw
}
\]
while the second contains the terms with coefficients $c_{(3,1,1)}, c_{(3,2,2)}$ and is implemented by the circuit
\[ \Qcircuit @C=1em @R=.7em {
	\lstick{q_0} & \qw & \qw & \ctrl{1} & \qw & \qw & \qw & \ctrl{1} & \qw & \qw & \qw \\
	\lstick{q_1} & \ctrl{1} & \gate{H} & \targ & \ctrl{1} & \gate{R_Z\left(-\frac{\xi \Delta t}{2\sqrt{2}}\right)} & \ctrl{1} & \targ & \gate{H} & \ctrl{1} & \qw \\
	\lstick{q_2} & \targ & \qw & \qw & \targ & \gate{R_Z\left(-\frac{\xi \Delta t}{2\sqrt{2}}\right)} & \targ & \qw & \qw & \targ & \qw
}
\]
\end{itemize}

The composition of all these pieces yields the total circuit that represents a single Trotter step evolution. After further simplification emerging form 
gate composition and available commutations, we obtain the circuit of Fig. \ref{fig:4sites_results_best}.

 \section[\appendixname~\thesection]{The circuit of Mach-Zender interferometer \label{secC1}}
 
\begin{figure}[t!]
	\centering
	\includegraphics[width=0.48\linewidth]{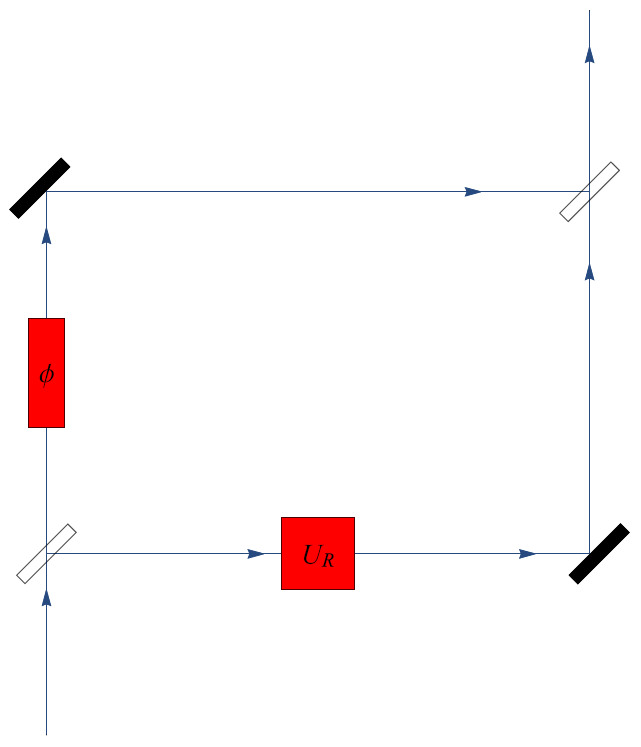}
	\caption{A Mach-Zender interferometer with mirrors and beam splitters represented by black and white rectangles respectively. The implementation of a control unitary is referred to red blocks.}
	\label{fig:mach_zender}
\end{figure}

The circuit presented in Section \ref{secCF} to evaluate correlation functions is based on the Mach-Zender interferometer \cite{vedral2000,vedral2003}. The setup is referred to Fig. \ref{fig:mach_zender}, where arrows identify input and output ports. White and black rectangles represent balanced beam splitters and mirrors respectively. These devices act on the ancilla $A$ degree of freedom $\mathscr{H}_A = \mathrm{span}\{  \ket{0}, \ket{1} \}$ related with wave packets moving along vertical and horizontal directions. We include internal degrees of freedom associated with our register $R$ according to the tensor product $\mathscr{H}_A \otimes \mathscr{H}_R$, such that the input state is $\varrho_{\mathrm{in}} = \ket{0}\bra{0} \otimes \varrho_R$, elaborated by mirrors $U_M = X \otimes \mathds{1}$, beam splitters $U_B = H \otimes \mathds{1}$ and a controlled gate
\begin{equation} \tag{E1}
    U = \begin{pmatrix}
    \e^{\ii \phi} & 0 \\
    0 & 0 
    \end{pmatrix} \otimes \mathds{1} + \begin{pmatrix}
    0 & 0 \\
    0 & 1 
    \end{pmatrix} \otimes U_R.
\end{equation}
The output state is defined by the transformation
\begin{equation} \tag{E2}
    \varrho_{\mathrm{in}} \rightarrow \varrho_{\mathrm{out}} =U_B U_M U U_B \varrho_{\mathrm{in}} U_B^\dagger U^\dagger U_M^\dagger U_B^\dagger,
\end{equation}
that yields the following expression
\begin{align} \label{out_mach_zender} \tag{E3}
    \varrho_{\mathrm{out}} = \frac{1}{4} & \left[ \begin{pmatrix}
    1 & 1 \\
    1 & 1 
    \end{pmatrix} \otimes U_R \varrho_R U_R^\dagger + \begin{pmatrix}
    1 & -1 \\
    -1 & 1 
    \end{pmatrix} \otimes \varrho_R \right. \\
    & \left. \ \ + \e^{\ii \phi} \begin{pmatrix}
    1 & 1 \\
    -1 & -1 
    \end{pmatrix} \otimes \varrho_R U_R^\dagger + \e^{-\ii \phi} \begin{pmatrix}
    1 & -1 \\
    1 & -1 
    \end{pmatrix} \otimes U_R \varrho_R \right]. \nonumber
\end{align}
The resulting intensity along the vertical direction is obtained by projecting the output state
\begin{equation} \tag{E4}
    \mathrm{Tr}_{AR} \{ (\ket{0}\bra{0} \otimes \mathds{1}) \varrho_{\mathrm{out}} \} = \frac{1}{2} \big( 1 + |\mathrm{Tr}_R\{ U_R \varrho_R \}| \cos(\phi-\mathrm{arg} (\mathrm{Tr}_R\{ U_R \varrho_R \})) \big),
\end{equation}
where we used $\mathrm{Tr}_R\{ \varrho_R U_R^\dagger \}=\mathrm{Tr}_R\{ U_R \varrho_R \}^*$ in order to introduce the visibility \cite{vedral2000} of the interference pattern $v = |\mathrm{Tr}_R\{ U_R \varrho_R \}|$. For a register in a pure state $\ket{\psi_0}$ the visibility reads $v = |\braket{\psi_0 |U_R| \psi_0}|$, while in a general mixed state $\varrho_R = \sum_k w_k \ket{k}\bra{k}$
\begin{equation}\tag{E5}
    v = \left|\sum_k w_k v_k \e^{\ii \phi_k}\right|,
\end{equation}
with $v_k = |\braket{k |U_R| k}|$ and $\phi_k = \mathrm{arg}(\braket{k |U_R| k})$ \cite{vedral2000}.

The diagonal elements of the Mach-Zender interferometer output in Fig. \ref{fig:mach_zender} are analogous to the ones of the output of the circuit \cite{ekert2002}
\[ \Qcircuit @C=1em @R=.7em {\lstick{\ket{0}} & \gate{H} & \gate{\phi} & \gate{X} & \ctrl{1} & \gate{H} & \meter \\
\lstick{\varrho_R}  & \qw & \qw & \qw & \gate{U_R} & \qw & \qw
}\]
yielding opposite sign for off-diagonal elements with respect to Eq. \eqref{out_mach_zender}
\begin{align}\tag{E6}
    \varrho_{\mathrm{out}} = \frac{1}{4} & \left[ \begin{pmatrix}
    1 & -1 \\
    -1 & 1 
    \end{pmatrix} \otimes U_R \varrho_R U_R^\dagger + \begin{pmatrix}
    1 & 1 \\
    1 & 1 
    \end{pmatrix} \otimes \varrho_R \right. \\
    & \left. \ \ + \e^{\ii \phi} \begin{pmatrix}
    1 & -1 \\
    1 & -1 
    \end{pmatrix} \otimes \varrho_R U_R^\dagger + \e^{-\ii \phi} \begin{pmatrix}
    1 & 1 \\
    -1 & -1 
    \end{pmatrix} \otimes U_R \varrho_R \right]. \nonumber
\end{align}
followed by the measurement
\begin{equation}\tag{E7}
    \mathrm{Tr}_{AR} \{ (Z \otimes \mathds{1}) \varrho_{\mathrm{out}} \} = \left\vert \mathrm{Tr}_{R} \{ U_R \varrho_R \} \right\vert \mathrm{cos}\left( \phi - \mathrm{arg}(\mathrm{Tr}_{R} \{ U_R \varrho_R \}) \right).
\end{equation}

 \section[\appendixname~\thesection]{T-REx and ZNE mitigation schemes \label{secD1}}
 
 In this Appendix we give a brief description of the two error mitigation techniques used in the paper.

The first method consists in T-REx, which focuses on readout errors \cite{trex}. An operator $V$ acts on a system of $N$ qubits, modifying its state as follows
\begin{equation} \label{Eq::circuit_op} \tag{F1}
    \varrho_{\mathrm{in}} \rightarrow \varrho_{\mathrm{out}} = V \varrho_{\mathrm{in}} V^\dagger,
\end{equation}
that we have to sample with respect to possible output strings $x \in \mathbb{Z}_2^N$ by means of a positive operator-valued measure (POVM) $E_x = \ket{x} \bra{x}$. We will assume that a noise map $A$ affects measurements, representing a $2^N \times 2^N$ left-stochastic matrix, which yields a noisy readout distribution $\widetilde{p} = A p$: the element $A_{x,y} = \braket{x | A |y}$ quantifies the probability of measuring $y$ instead of $x$, defining a noisy POVM $\widetilde{E}_x = \sum_y A_{x,y} \ket{y} \bra{y}$.

Given a string $s \in \mathbb{Z}_2^N$, it is possible to define
\begin{equation} \tag{F2}
    Z_s = \sum_x (-1)^{\braket{s,x}} \ket{x}\bra{x}, \quad X_s = \sum_x \ket{x+s}\bra{x} = \sum_x \ket{x}\bra{x+s}=X_s^\dagger,
\end{equation}
involved in the targeted estimation of
\begin{equation} \tag{F3}
    \braket{Z_s} = \mathrm{Tr}\left\{ Z_s \varrho_{\mathrm{out}} \right\} = \sum_x (-1)^{\braket{s,x}} \mathrm{Tr}\left\{ E_x \varrho_{\mathrm{out}} \right\}
\end{equation}
whose unmitigated noisy version is $\braket{\widetilde{Z}_s} = \sum_x (-1)^{\braket{s,x}} \mathrm{Tr}\left\{ \widetilde{E}_x \varrho_{\mathrm{out}} \right\}$. The mitigation of this readout error is achieved by applying random bit flips before and after the noisy measurements, as expressed by the twirled noise map and associated POVM
\begin{equation} \tag{F4}
    A^* = \frac{1}{2^N} \sum_s X_s A X_s^\dagger, \qquad \widetilde{E}_x^* = \sum_y A_{x,y}^* \ket{y} \bra{y}.
\end{equation}
The eigensystem of this map is obtained by exploiting the definition of the eigenvectors $\ket{v_w} = \sum_x (-1)^{\braket{w,x}} \ket{x}$, such that the eigenvalue equation is deduced
\begin{align} \tag{F5}
    A^* \ket{v_w} = & \ \frac{1}{2^N} \sum_{s,x} \sum_{a,b} (-1)^{\braket{w,x}} A_{a,b} \ket{a+s} \braket{b+s | x} \\ 
    =& \ \frac{1}{2^N} \sum_{x} \sum_{a,b} (-1)^{\braket{w,x+a+b}} A_{a,b} \ket{x} = \lambda_w \ket{v_w}, \nonumber
\end{align}
yielding $\lambda_w = \frac{1}{2^N} \sum_{a,b} (-1)^{\braket{w,a+b}} A_{a,b}$. The twirled noise expectation reads
\begin{align} \tag{F6}
    \braket{\widetilde{Z}_w^*} = & \ \sum_x (-1)^{\braket{w,x}} \mathrm{Tr}\left\{\widetilde{E}_x^* \varrho_{\mathrm{out}} \right\} = \sum_{x,y} (-1)^{\braket{w,x}} \braket{x | A^* | y } \mathrm{Tr}\left\{ \ket{y} \bra{y} \varrho_{\mathrm{out}} \right\} \\
    = & \ \sum_{y} \braket{v_w | A^* | y } \mathrm{Tr}\left\{ \ket{y} \bra{y} \varrho_{\mathrm{out}} \right\} = \lambda_w \braket{Z_w}, \nonumber
\end{align}
that for the state $\varrho_{\mathrm{out}} = \ket{0}\bra{0}$ allows us to estimate $\braket{\widetilde{Z}_w^*} = \lambda_w$, since $\braket{Z_w} = 1$. The same quantity is used to obtain noise mitigation for other values of $\varrho_{\mathrm{out}}$.

The second technique is referred to ZNE, adopting a noise scaling in order to extrapolate noiseless expectation values \cite{zne}. In practice a ``de-optimizing'' procedure is applied by increasing the circuit depth. We assume that the action of a circuit, as expressed in Eq. \eqref{Eq::circuit_op}, is composed of $d$ unitary layers
\begin{equation} \tag{F7}
    V = L_d L_{d-1} \dots L_2 L_1,
\end{equation}
such that the depth is equal to $d$. A first step aimed at noise scaling consists in unitary folding, which replaces the unitary circuit by
\begin{equation} \tag{F8}
    V \rightarrow V(V^\dagger V)^n,
\end{equation}
where $n$ is a positive integer leading to a new depth $(2n + 1) d$, with no logical effect since $V^\dagger V = \mathds{1}$. In order to improve the scaling resolution it is possible to exploit also a partial folding concerning the last $s$ layers of the circuit
\begin{equation} \tag{F9}
    V \rightarrow V(V^\dagger V)^n L_d^\dagger L_{d-1}^\dagger \dots L_{d-s}^\dagger L_{d-s} \dots L_{d-1} L_d,
\end{equation}
whose depth is $(2n+1)d + 2s$. The depth stretching $d \rightarrow \lambda d$ is ruled by a scale resolution $2/d$
\begin{equation} \tag{F10}
    \lambda = 1 + \frac{2 k}{d},
\end{equation}
with $k = 1,2,\dots,nd+s$. We assume a linear, a polynomial or an exponential dependence of observables expectation values with respect to the scaling parameter $\braket{O(\lambda)}$, where $\braket{O(1)}$ is the natural noise and we extrapolate the noiseless case $\braket{O(0)}$.


\bibliographystyle{apsrev4-2}

\bibliography{bibliography.bib}

\end{document}